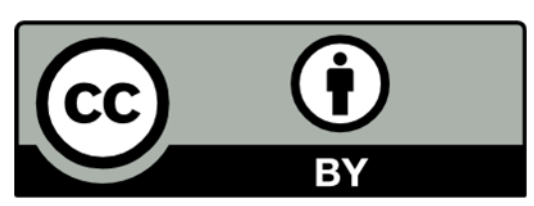

# Spin-orbit coupling by design in quantum state engineering of atomically defined quantum dots

Hermann Osterhage[1]†, Julian H. Strik[1]†, Ivan Ado[1], Anna M. H. Krieg[1], Daniel Wegner[1], Mikhail Titov[1], Alexander A. Khajetoorians[1*]

[1]*Institute for Molecules and Materials, Radboud University; Nijmegen, The Netherlands*

*Corresponding author: a.khajetoorians@science.ru.nl; †equally contributing first authors

**Tuning spin-orbit coupling is essential in controlling both spin and charge in confined semiconductor nanostructures, yet it is rarely a truly controllable parameter. Here, we show control over the spin-orbit Hamiltonian in quantum dots and the resulting quantum states by tailoring the confinement potential with atomic-scale precision. Using scanning tunnelling microscopy and spectroscopy, we pattern individual Cs ions into designer quantum dot structures on the surface of indium antimonide, in which electrons from a two-dimensional electron gas are confined with chosen in-plane electric-field gradients. We then quantify the atomic level structure, both spatially resolving the orbital character of the electronic states and their magnetic-field evolution. We demonstrate that the level structure, including the induced zero-field splitting, can be tailored by the designed geometry of the local electric fields. These effects can be described using a Hamiltonian that allows consistent treatment of the confinement-induced spin-orbit coupling beyond the conventional Bychkov-Rashba description. This Hamiltonian is derived from a multiband k·p model and takes the energy dependence of the relevant physical parameters into account. Such precise control of spin-**

**orbit coupling in semiconductor quantum dots is relevant to quantum and spintronic technologies.**

Spin-orbit coupling (SOC) links the angular motion of an electron to its spin, providing a mechanism to link the charge and spin degrees of freedom in solids. Tuning SOC in materials plays a central role in the design and functionality in spintronic techonologies[1], engineering topological phases[2,3] and realizing semiconductor qubits[4-8]. In semiconductor nanostructures, SOC is most prominently tuned electrostatically, using the Rashba effect. The Rashba effect is based on using an electric field oriented along the normal of a two-dimensional electron gas (2DEG), converting momentum into an effective in-plane magnetic field[9,10]. Gate control and engineering of Rashba-type SOC have been demonstrated in transport, quantum wells and oxide interfaces[11-13]. Electrostatic SOC tuning via the Rashba effect has been explored mainly in one- and two-dimensional systems with extended interfaces that generate a normal electric field.

Realizing tunable SOC in zero-dimensional electronically confined structures, where atomic level structures are engineered, however, has been challenging. Experimentally, the presence of SOC has been seen in self-assembled InAs quantum dots (QDs)[6], resulting from potential gradients present at the interfaces. The interplay between Rashba SOC and confinement geometry in QDs has been studied theoretically, considering concepts like intrinsic spin mixing, *g*-tensor anisotropy, exact solutions and coupled-dot extensions[14-21]. Experimentally, the challenge to tune SOC in QD structures is in engineering the necessary electric fields to break symmetry: in most QDs the lateral fields generated at a particular interface are weak and less controllable than in vertically confined structures[22,23]. This limits the tunability of the SOC in conventional quantum dots mainly to the Rashba contribution, often treated as a small correction to the defined atomic states. These states are typically described by quantum states of the lateral confinement, so called Fock-Darwin states[24-26], while SOC contributions from the lateral fields are often neglected. To date it is still challenging to fabricate QDs with a designed multiplet structure with chosen SOC.

Quantum dots on the surface of semiconductors can be created with atomic length scales and precision, using scanning tunneling microscopy (STM)[27-29]. Based on this idea, it was shown that coupled QDs with multi-orbital character can be created by locally sculpting the electrostatic potential in a dilute 2DEG using the tip of the STM[30]. Here, we design the electrostatic potential by locally patterning ions on the surface of a dilute two-dimensional electron gas, to pattern QDs with an appreciable SOC (Fig. 1a). We start with the cleaved surface of undoped InSb(110), which features no surface states in its native band gap. Individual Cs adsorbates act as ionic dopants that bend the conduction band downward leading to the creation of a two-dimensional electron gas near the surface[31], with carrier densities typically on the order of $6 \times 10^{15}\ m^{-2}$ to $3 \times 10^{16}\ m^{-2}$ [32,33]. By patterning Cs atoms into designer arrays in regions with an absence of Cs atoms, we create designed electrostatic potential gradients with nanometer-scale variation. In this way, 0D a confinement potential with a chosen shape, depth, and in-plane potential gradients can be crafted with nanometer precision. Using scanning tunnelling spectroscopy (STS) in an applied magnetic field, we quantify the excitation spectrum of the quantum states of both isotropic and anisotropic QDs, as well as image the spatially confined wavefunctions and SOC-induced zero-field splittings. We quantitatively reproduce the multiplet structure and its magnetic field response, using a multiband theory in the k·p approach[34]. We employ a novel method to derive the energy-dependent effective Hamiltonian of the conduction band. This allows us to quantify how strong confinement modifies the effective parameters (effective mass, $g$-factor and SOC constants). This enables precise design of the multiplet structure of QDs, namely atomic precision of the relevant confinement gradients and predictable control of SOC-induced energy splittings and magnetic response in semiconductor QDs.

The combination of low electron density in the underlying 2DEG and higher local density of Cs atoms leads to a local 0D confinement potential well with a smooth profile, as sketched in Fig. 1b,c. Depending on the local Cs density, the depth of the attractive potential can be tuned. We consider the Hamiltonian $H_{00} = \frac{\hbar^2}{2m^*}\boldsymbol{k}^2 + V$ for $B = 0$ and no SOC with $\boldsymbol{k} = -i\boldsymbol{\nabla}$ and the total confinement potential (see Supplementary Figs. S1-S3)

$$V = vz + \kappa_x x^2 + \kappa_y y^2. \quad (1)$$

This corresponds to the electrostatic potential close to the center of a charged disk, or more generally a charged ellipse. $V$ is the sum of a vertical confinement defined by a triangular well in z[33], and a lateral confinement term defined by a generalized harmonic potential with anisotropic components $\kappa_i = m^*\omega_i^2/2$, with effective mass $m^*$ and oscillator frequency $\omega_i$. We focus on the lowest quantum level of the triangular vertical confinement. For isotropic lateral confinement in the disk geometry ($\kappa_x = \kappa_y$), the eigenstates ($n$, $l$, $s$) of the lateral part of the Hamiltonian $H_{00,\parallel} = -\frac{\hbar^2}{2m^*}(\nabla_x^2 + \nabla_y^2) + \kappa_x x^2 + \kappa_y y^2$ can be described by polar quantum numbers ($n$, $l$), where $n$ is the radial quantum number, $l$ describes the angular momentum, and a spin projection $s$ (Fig. 1d). Anisotropy can also be induced in the spatially varying potential by creating an anisotropic, elliptic distribution of Cs atoms (Fig. 1c). In this case, radial symmetry is broken, and the eigenstates can be described by cartesian quantum numbers ($n_x$, $n_y$), as well as the additional spin quantum number $s$. The influence of an out-of-plane magnetic field on electronic states in a parabolic confinement potential is conventionally described by so-called Fock-Darwin states[24]. In this paper we will show deviations of the energy spectrum of atomic-scale QDs from the Fock-Darwin model without SOC and explain the observations by including confinement-induced Rashba[15] and lateral SOC contributions.

In the presence of a strong potential gradient, an additional contribution to the electronic spectrum arises due to SOC, namely $H_{\mathrm{SOC}} = \lambda_{\mathrm{SOC}}[\boldsymbol{\nabla} V \times \boldsymbol{k}]\boldsymbol{\sigma}$, where $V$ is the confinement potential, and $\boldsymbol{k}$, $\boldsymbol{\sigma}$ represent the electron momentum and the Pauli matrices, respectively. In the limit of a 2D plane, this yields the expected Rashba term. The introduction of SOC leads to mixing of spin and orbital angular momentum. For $\lambda_{\mathrm{SOC}} \neq 0$, the fourfold degeneracy of states with quantum numbers $(n, \pm l, \pm 1/2)$ is lifted, and twofold degenerate states emerge with angular momentum and spin being parallel or antiparallel (see Fig. 1e). This breaking of the degeneracy in the multiplet structure at zero field, due to the SOC, is similar to a zero-field splitting induced by a crystal field[35-38]. Recently, it was found that the potential $V$ leads to an enhancement of $\lambda_{\mathrm{SOC}}$ in InSb by $\lambda'$ due to interband mixing[39], and thus to an enhanced SOC-induced splitting (see Supplementary S1)[40]. The different states in the multiplet can be distinguished by their dependence on magnetic field as shown in Fig. 1f due to the Zeeman term, $H_{\mathrm{Z}} = \frac{g^*}{2}\mu_{\mathrm{B}}\boldsymbol{B}\boldsymbol{\sigma}$, with an effective $g$-factor $g^*$, and due to the B-dependence of the momentum: $\boldsymbol{k} = -i\boldsymbol{\nabla} - e\boldsymbol{A}$ with the vector potential $\boldsymbol{A} = \frac{1}{2}\boldsymbol{B} \times \boldsymbol{r}$.

We subsequently created individual atomic-scale QDs, as described above, with a designed confinement potential, to probe both the resultant electronic states and the role of SOC. An evenly spaced square array with octagonal edges is shown in Fig. 2a, where the interatomic distance between Cs atoms is 3.2 nm. This structure and those shown later are isolated from the disordered 2DEG on the surface, by moving away any unwanted Cs atoms in the vicinity of the structure within a distance of 25 nm. This leads to a depletion region where the local electronic structure is sufficiently decoupled from the surrounding 2DEG. STS was acquired at various points in the upper left quadrant of the QD, as indicated by the "x" markers in Fig. 2a. STS revealed six discrete peaks, where the intensity of each peak depends on the spatial location where it was measured in the QD (Fig. 2b). These peaks result from the local confinement potential induced by the Cs atoms. To infer the confined wavefunctions, we measured the spatially resolved differential conductance ($\mathrm{d}I/\mathrm{d}V$) at these peak energies as shown in Fig. 2c. The four states lowest in energy showed radial symmetry, suggesting an isotropic confinement potential, corresponding to a charge distribution in a disk-like geometry on the surface. The last two measured states show a more complex pattern that breaks radial symmetry. They indicate a four-lobed structure, at different energies, but are the same pattern up to a 45° rotation. We note that the structure is not perfectly circular, which may lead to small anisotropic corrections near the top of the well created by the lateral confinement potential.

The inferred wavefunctions and the responsible confinement potential can be modeled in the simplest approximation by an isotropic 2D harmonic oscillator potential, neglecting the SOC. In this case, the eigenstates are described by quantum numbers $(n, l, s)$. We calculated the spatially dependent wavefunction $|\Psi|^2$, first considering an isotropic potential (see Supplementary S1)[40]. Comparison to the eigenstates of $H_{00}$ with $\hbar\omega_0 = 70$ meV, and $m^* = 0.02\, m_\mathrm{e}$, depicted in Fig. 2d, showed a resemblance of the charge distribution in the QD to the model of parabolic confinement. While the spatially dependent wavefunctions can be well captured in this simple approximation, the states with $(n,l,s) = (0,\pm2, \pm1/2)$ and $(1,0, \pm1/2)$ are predicted to be fully degenerate in the Fock-Darwin model without SOC (see Fig. 1d). Deviating from this model, we observe non-degeneracy of states at $V_S$ = -77 mV, -68 mV, and -56 mV. The lowering of the (1,0, ±1/2) levels relative to (0,±2, ±1/2) could potentially be due to anharmonicity of the confinement potential combined with their stronger localization in the center of the well. The splitting within the (0,±2, ±1/2) multiplet of (11±1) meV found in Fig. 2b we is attributed majorly to SOC. For comparison, in a quantum dot of circular symmetry with an even more isotropic confinement potential this splitting was still 9 meV (see Supplementary Fig. S8)[40]. Another striking difference compared to the model is that the peaks marked yellow and green in Fig. 2b are found at different energies experimentally ($V_S$ = -132 mV and -125 mV) but show the same charge density distribution. In the absence of spin-orbit coupling, these states are degenerate with quantum numbers $(n, l, s) = (0, \pm1, \pm1/2)$. The splitting in energy in the absence of magnetic field between these states is a further indication of SOC, as depicted in Fig. 1e. Above -50 mV we observed broader features in the spectra, probably due to the onset of additional 2DEG subbands that complicate the interpretation of these states. Because of this we limit the discussion in this work to the states confined at $V_S < -50$ mV.

We can additionally introduce potential anisotropy by changing the spatial distribution of Cs atoms. In order to probe the effect of such an anisotropy on the tailored electronic states and its SOC, we constructed an anisotropic QD structure, as shown in Fig. 3. In Fig. 3a, we show an elongated QD constructed from 36 Cs atoms, with the same interatomic distance of 3.2 nm as for the previously discussed QD in Fig. 2. STS was acquired at various points in the upper left quadrant of the QD, as indicated by the "x" in Fig. 3a. The resultant spectra acquired at various positions inside this QD are shown in Fig. 3b with colored arrows marking the first five discrete states. The peaks appear at different energies than in the isotropic case, and the energy spacing between levels is no longer equidistant nor equivalent to the previous case. Differential conductance maps in Fig. 3c revealed confinement states with broken radial symmetry. The ground state at -205 mV is elliptic and elongated along the horizontal ($x$) direction. The states at -149 mV and -124 mV are dumbbell-shaped with nodal lines along $y$- and $x$- direction, respectively. The next state at -93 mV shows two nodal lines along $y$ and the state at -74 mV is clover-like with one nodal line each along $x$ and $y$. To ascertain the role of SOC, we first modeled the lateral confinement potential without SOC using two linearly separable 1D harmonic potentials. In this case, the problem can be parameterized by the quantum numbers $n_x$, $n_y$, and spin $s$. In the anisotropic case, due to the loss of radial symmetry, a cartesian representation is used and angular momentum $l$ is no longer a good quantum number. Calculations of the expected wavefunctions coincide well with the measured charge density. Fig. 3d shows the charge distribution of the QD states corresponding to eigenfunctions of $H_{00}$ with $\hbar\omega_x = 57$ meV, $\hbar\omega_{xy} = 81$ meV, and $m^* = 0.02\, m_e$. This illustrates that the lateral confinement strength can be tuned in-plane, depending on the Cs layout. The influence of SOC on the level structure in this case is more subtle because all SOC-related terms are non-diagonal in the basis $(n_x,n_y,s)$. For example, the states $(n_x,n_y,s) = (0,1,\pm1/2)$ and $(1,0,\pm1/2)$ are non-degenerate because of the anisotropic confinement. SOC will mix these states and effectively shift the levels relative to each other. However, this influence of SOC on the zero-field spectrum is expected to be weak compared to the splitting due to anisotropic confinement.

To further quantify the multiplet structure and the role of SOC, for both QD structures (disk and ellipse in Figs. 2-3), we investigated the response of the eigenstates of each QD to an external out-of-plane magnetic field. For the isotropic QD shown in Fig. 2, STS was acquired at the six sites marked in Fig. 2a in intervals of 0.1 T, and the resultant STS was averaged spatially and is shown in Fig. 4a. We can identify the states marked with colored arrows in Fig. 2a at $B$ = 0 T and track their evolution in increasing magnetic field individually. We label the states based on their symmetry at zero field as part of a Fock-Darwin multiplet ($n$,$l$,$s$). The state (0,0,±1/2) (red) remains relatively unchanged in magnetic field, but ultimately splits into two peaks in high magnetic field. The states (0,±1,±1/2) (green and yellow) are non-degenerate in zero field, due to SOC, and fan out into three distinguishable branches as the magnetic field increases with the branch lowest in energy showing the highest intensity. The states (1,0,±1/2) (light blue), and (0,±2,±1/2) (dark blue and purple) each split into two peaks with different slopes, reflecting the relative alignment of angular momentum $l$ and spin $s$ as sketched in Fig. 1e,f. Here, the role of SOC is revealed in the zero-field splitting between the states marked green and yellow and between dark blue and purple, differing in ($l$+$s$), and in the different magnitude of the splitting of these states in magnetic field.

For the anisotropic QD, shown in Fig. 3, the magnetic field evolution is shown in Fig. 4c. STS was averaged over the eight sites marked in Fig. 3a. The state ($n_x$,$n_y$,$s$) = (0,0,±1/2) (red) remains relatively unchanged in magnetic field but ultimately splits into two peaks in high magnetic field, similar to the isotropic case. The state (1,0,±1/2) (yellow) with one node in the $x$-direction moves to lower energies in magnetic field and does not show any visible splitting up to 9 T. On the other hand, the state (0,1,±1/2) (green) with a node in the $y$-direction splits even more strongly in magnetic field than the lowest energy state and curves upwards towards higher energies. The state (2,0,±1/2) with two nodes along the $x$-direction again curves downwards and does not split visibly in magnetic field. Lastly, the state (1,1,±1/2) (purple) shifts non-monotonously and splits visibly in magnetic field. The data therefore shows alternating magnetic-field-induced splittings across successive anisotropic states. These alternating splittings cannot be explained by the energy dependence of the substrate's $g$-factor due to the band structure, as those effects are monotonous in energy[34,41]. Instead, the alternating splitting is correlated with the nodes of the wavefunction being found along the weakly confined direction (weak splitting) or along the strongly confined direction (larger splitting), see also Supplementary Figs. S9 and S10.

In order to quantify the expected influence on SOC on the electronic states of both QDs, we used an analytical model. Here, we assume the confinement potential shape in equation (1), extract the confinement strength from STS data and calculate the inferred energy spectrum including the effect of Rashba-type and lateral SOC. The model Hamiltonian used is:

$$H = \frac{1}{2m^*}(-i\hbar\boldsymbol{\nabla} - e\boldsymbol{A})^2 + V + H_{\mathrm{SOC}} + H_{\mathrm{Z}}. \tag{2}$$

The energy dependence of the effective mass $m^*$ and the $g$-factor $g^*$ are considered via the eight-band Kane model[34,39,40,42]. $\lambda_{\mathrm{SOC}}$ is a material- and energy-dependent SOC parameter. $H_{\mathrm{SOC}}$ depends critically on the gradient of the confinement potential $\boldsymbol{\nabla}V$. We approximated the confinement potential $V$ as a sum of a triangular well that depends on $z$[33], leading to Rashba SOC[6,16,20] and a parabolic well in the $x$-$y$ plane (see equation (1)), adding additional SOC contributions to the Hamiltonian. The lateral confinement strength $\kappa_{\mathrm{i}}$ was extracted from the spatial decay of the ground state at $B$ = 0 T (see Supplementary Figs. S6 and S7). The splitting of the ground state in $B$ was found to mainly originate from the Zeeman term $H_{\mathrm{Z}}$, which allowed the determination of the vertical confinement strength $\nu$ from the energy dependence of the $g$-factor according to the Kane model[40] (see Supplementary Figs. S4 and S5). The advantage of this approach is that there is no *ab-initio* determination of parameters and that the model is not fit directly to the experimental spectrum by optimization of parameters. Instead, we assume only the shape of the potential, extract the confinement strengths from the $g$-factor and spatial decay of the ground state, and then compare the model calculation to the experiment.

We calculated the magnetic field-dependent energy spectra by treating $H_{SOC}$ as a perturbation to the Fock-Darwin eigenstates in the absence of SOC[24]. The energy dependence of $\lambda_{SOC}$, $m^*$, and $g^*$ are considered self-consistently as these parameters are sensitive to energy shifts and thus implicitly depend on the confinement strengths. For isotropic lateral confinement, the Fock-Darwin energies are $E_{0,\parallel}(n,l,s) = (2n+|l|+1)\hbar\Omega - \frac{\hbar\omega_c}{2}l + g^*\mu_B Bs$ with radial quantum number $n = 0,1,2,...$, angular quantum number $l = 0,\pm1,\pm2,....$, spin quantum number $s = \pm1/2$, the cyclotron frequency $\omega_c = -\frac{|e|B}{m^*}$, and the characteristic frequency $\Omega = \sqrt{\omega_0^2 + \omega_c^2/4}$. The calculated level diagram of the model Hamiltonian in equation (1) is shown in Fig. 4b for the case of isotropic lateral confinement ($\kappa_x = \kappa_y = 1.06\times10^{15}$ eV/m$^2$), and $\nu = 5.4\times10^7$ eV/m, extracted from the ground state energy value (see Supplementary S2)[40]. Here, we label each Kramer's pair with its total angular momentum ($l$+$s$). Despite the simplicity of the model with only two parameters, the model reproduced the main features of the experimentally observed multiplet structure. The comparably small splitting in magnetic field of states $(n,l,s) = (0,0,\pm1/2)$ and $(1,0,\pm1/2)$ matched the observations for corresponding states marked red and light blue in Fig. 4a. Likewise, the stronger splitting of the other calculated Kramer's pairs matched the experimental trends. The zero-field splitting of states with $(n,l,s) = (0,\pm1,\pm1/2)$ into Kramer's pairs with $(l+s) = (\pm1/2)$ and $(\pm3/2)$ was calculated as 7.3 meV, matching the (8±1) meV found experimentally. We note that this splitting would be underestimated by a factor of four without the recently found contribution $\lambda'$ to the SOC[39,40]. Here, 5.8 meV stems from SOC induced by the lateral confinement $\kappa_{\mathrm{i}}$ and 1.5 meV are due to the Rashba term. This shows that the lateral potential gradient due to the lower dimensional confinement potential significantly modulates the SOC, on the same order of magnitude as the Rashba effect. With 10.3 meV, the model also reproduces the zero-field splitting measured for states $(n, l, s) = (0, \pm 2, \pm 1/2)$ (11±1 meV in the experiment).

In the anisotropic case, the circular symmetry is broken and degeneracies are lifted. Consequently, the angular momentum operator does not commute with the Hamiltonian and we treat the term $-\frac{\omega_c}{2}\boldsymbol{L_z}$ perturbatively[40]. The eigenenergies of the unperturbed problem are $E_{0,\parallel}(n_{\mathrm{x}}, n_{\mathrm{y}}, s) = (n_{\mathrm{x}} + 1/2)\hbar\Omega_{\mathrm{x}} + (n_y + 1/2)\hbar\Omega_y + g^*\mu_B Bs$, with quantum numbers $n_i = 0,1,2, \ldots$ and separate oscillator frequencies $\Omega_{\mathrm{x}}$ and $\Omega_{\mathrm{x}}$. The calculated magnetic field-dependent energy spectrum of the anisotropic QD with $\kappa_{\mathrm{x}} = 0.94\times10^{15}$ eV/m$^2$, $\kappa_{\mathrm{y}} = 1.85\times10^{15}$ eV/m$^2$, and $\nu = 5.1\times10^{7}$ eV/m is shown in Fig. 4d. The calculations reproduced the experimental observation of alternating magnitude of *B*-field induced energy splitting, mimicking a state-dependent renormalization of the *g*-factor. This effect cannot be explained by the conventional description of Fock-Darwin states with purely Rashba-type SOC[15] but is a fingerprint of the lateral potential gradient. To understand this behavior qualitatively, we analyzed the matrix elements of the perturbative terms. The magnetic field and the SOC due to the in-plane potential are mixing the states of the unperturbed problem (1,0,1/2) with (0,1,1/2) and (1,0,-1/2) with (0,1,-1/2). Assuming the perturbation and Zeeman energy to be small compared to $\hbar\omega_y - \hbar\omega_x$ yields an effective reduction of the *B*-induced splitting of (1,0,±1/2) by $\Delta E = m^*\lambda_{\mathrm{SOC}}\omega_{\mathrm{c}}(\omega_{\mathrm{x}} + \omega_{\mathrm{y}})^2/2(\omega_{\mathrm{y}} - \omega_{\mathrm{x}})$ and an enhancement of the *B*-induced splitting of (0,1,±1/2) by the same amount[40]. Similarly, the splitting of states (2,0,±1/2) is reduced due to coupling to states (1,1,±1/2). These reductions and enhancements originate from the interplay of the term $-\frac{\omega_c}{2}\boldsymbol{L_z}$ and the SOC due to lateral confinement[40]. The other SOC terms in the Hamiltonian have weaker effects and do not lead to alternating contributions to the magnetic field splitting.

## Conclusion

Here, we show that SOC can be controlled in atomic-scale QDs, by tailoring electrostatic potentials with nanoscale variation. We pattern the electrostatic potentials by arranging individual Cs ions on the surface of a dilute 2DEG, leading to appreciable confinement, and quantify their multiplet structure by measuring the magnetic field response of the eigenstates and their spatial dependence. Atomically patterned electrostatic confinement in InSb quantum dots generates lateral potential-gradient SOC terms that are comparable to, and distinguishable from, conventional vertical Rashba SOC. These terms can be inferred from zero-field splittings and state-dependent magnetic field evolution of spatially resolved confined states. We consider two cases: (i) isotropic QDs, which feature large zero-field energy splittings within a multiplet structure, beyond the Fock–Darwin picture[24,43]; (ii) anisotropic QDs which show additional SOC-induced features, defined by a magnetic field evolution which is correlated with the nodal structure of the wavefunctions. We adopt a predictive theoretical model using a generalized Kane model, enabling tunable design of the resulting quantum states. We show that quantitative modelling of strongly confined states in narrow-gap semiconductors cannot be obtained by simply appending a Rashba term to a fixed-parameter effective-mass model[9,10,20]. The energy dependence of material parameters and the influence of the confinement potential on the SOC Hamiltonian have to be considered to precisely determine the electronic structure. The combination of atomic-scale electrostatic design and a gauge-consistent multiband reduction offers a route to Hamiltonian design in semiconductor nanostructures. Such laterally induced SOC is difficult to realize in conventional epitaxial semiconductor QDs, where lithographic lateral confinement offers less control than the typically stronger vertical confinement[22-24,43]. In nanowires, strong confinement can be realized in only two dimensions[44,45]. Our approach offers a design principle towards quantum-state engineering and magnetic response for spin-orbit-enabled qubit control[4,5,44]. Extending atomic patterning from single dots to coupled-dot arrays could enable lattices in which the local spin-orbit Hamiltonian is programmed by geometry and may differ from site to site, providing a new avenue for quantum simulation[30] and for engineered topological or spin-transport functionality in narrow-gap materials.

**Methods (including separate data and code availability statements)**

**Tip and sample preparation**

Electrochemically etched W tips were cleaned by electron-beam heating and were covered with Au by means of dipping into a Au(111) single crystal surface. Undoped InSb single crystals were mounted on metallic sample plates and contacted on the side with conductive epoxy glue to create an electrical contact to all layers of the crystal. The InSb crystals were then cleaved in UHV and precooled to temperatures below 30 K before evaporating Cs from a $Cs_2CrO_4$ dispenser (SAES Getters) onto the InSb(110) surface. We used samples with a typical coverage of 1-2×$10^{12}$ Cs atoms per $cm^2$, determined by counting individual Cs atoms in STM images.

**STM/STS measurements**

The experiments were performed under ultra-high vacuum (UHV) conditions in a home-built STM at a temperature of $T = 7$ K. STM topography images were acquired in constant-current mode with typical tunneling parameters of $V_S = 300$ mV and $I_t = 1$ pA. Atomic QDs were assembled by lateral atom manipulation with typical tunneling parameters in the range of $V_S = 12$ - 30 mV and $I_t = 20$ - 300 pA. Tunneling spectra were acquired with lock-in technique and open feedback loop after stabilizing at $V_S = V_{stab}$ and $I_t = I_{stab}$ with the modulation voltage set to $V_{mod} = 0.5$ mV rms at $f =$ 871 Hz. d$I$/d$V$ maps shown in Fig. 2 and 3 were obtained by using a multi-pass method where first a constant-current contour was recorded with $V_S$ = 300 mV and $I_t$ = 5 pA. This contour was subsequently followed with open feedback loop and the d$I$/d$V$ signal was registered at a particular voltage set to a peak position in the tunneling spectra. The signal strength was controlled by adding a constant z offset $z_{offset}$ on the order of 100 pm per voltage of interest. All data has been processed using MATLAB and Gwyddion software. Topographic maps were flattened with standard plane subtraction methods.

**Model calculations**

For the model calculations, corrections to the spin-dependent Fock-Darwin energies due to SOC and due to the magnetic vector potential were calculated perturbatively for the 12 lowest energy states of the Fock-Darwin problem. Here, the energy dependence of the physical parameters $\lambda_{SOC}$, $m^*$, and $g^*$ was taken into account self-consistently by means of an effective energy-dependent conduction-band Hamiltonian derived from the 8-band Kane model[34]. The recently found enhancement of $\lambda_{SOC}$ in InSb by $\lambda'$[40] due to interband mixing[39] provides the largest contributions to the observed SOC splittings.

**Data availability:** The data supporting the findings of this study are available upon reasonable requests. Source data are provided with this paper.

**Code availability**: The codes that support the findings of this study are available from the corresponding author upon reasonable request.

**Acknowledgements** The authors would like to thank Malte Rösner and Mikhail Katsnelson for scientific discussions. This publication is part of the project "What can we 'learn' with atoms?" (with project number VI.C.212.007) of the research program VICI which is (partly) financed by the Dutch Research Council (NWO). A.A.K. and J.H.S. acknowledge the research program "Materials for the Quantum Age" (QuMat) for financial support. A.M.H.K. and H.O. acknowledge funding from Volkswagen foundation (with project number 0071131). This program (registration number 024.005.006) is part of the Gravitation program financed by the Dutch Ministry of Education, Culture and Science (OCW).

**Author contributions** H.O., J.H.S, and A.M.H.K. performed the experiments. H.O. and J.H.S. performed the experimental analysis. All authors participated in the iterative discussions about the experimental data, its analysis and the scientific conclusions. I.A. and M.T. developed the k·p methods and performed the calculations. H.O., J.H.S., I.A., and M.T. made the comparisons between experiment and simulations, with all authors participating in the interpretation of this comparison. H.O., J.H.S., and A.A.K. designed the experiments. H.O., A.A.K., I.A., and M.T. primarily wrote the manuscript, while all authors provided input during its development.

**Competing interests:** The authors declare no competing interests.

**Supplementary Materials**

Supplementary Discussion S1- S3

Supplementary Figs. S1- S10

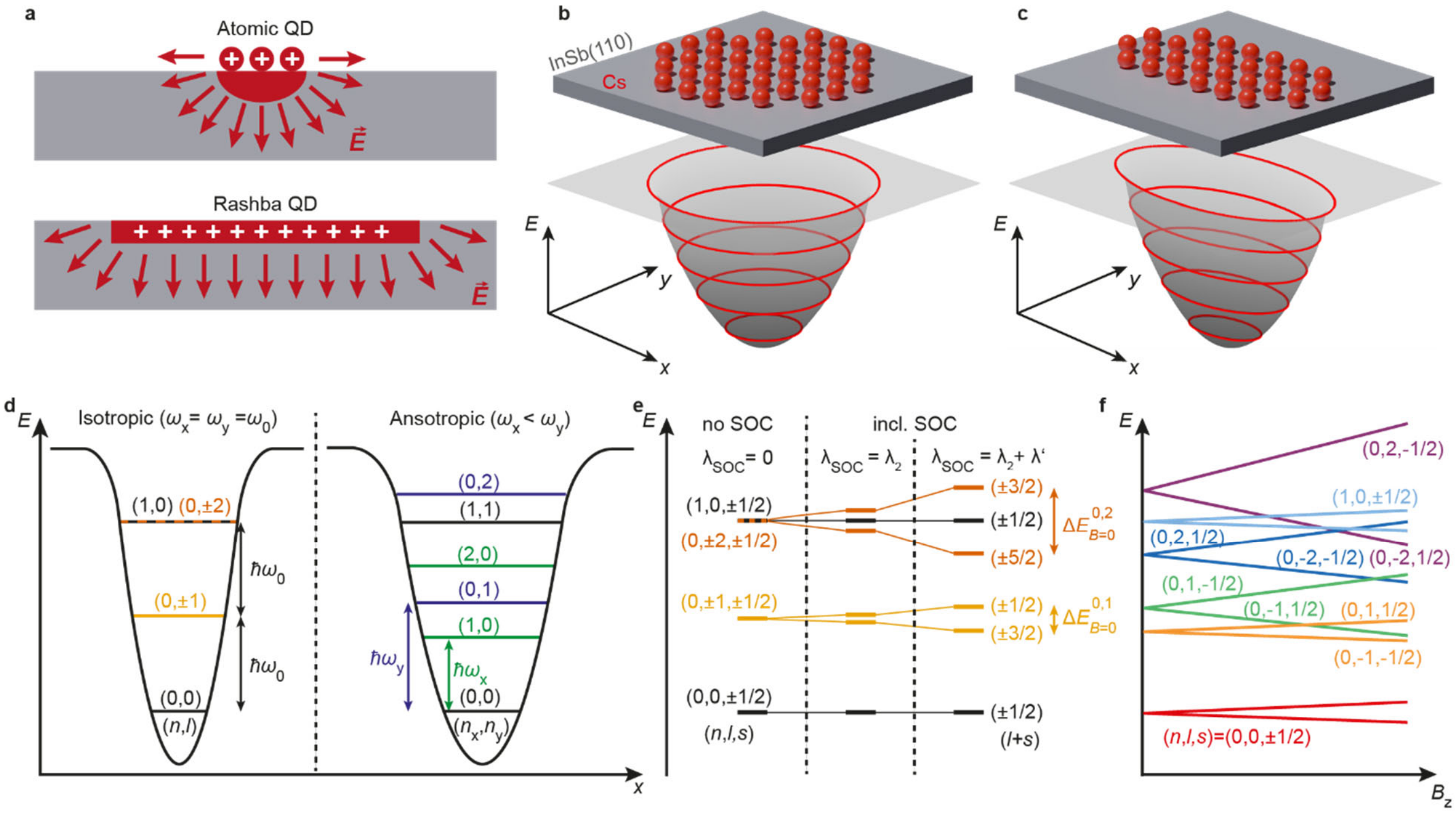

**Fig 1 | 0D confinement potential well created by Cs atoms on InSb(110) and the influence of SOC on the energy spectrum.**

**(a)** Comparison of electric fields radiating from an atomic QD formed by a few charged atoms with similar lateral and vertical electric field components (top) and a quasi-2D Rashba QD with mostly vertical electric field components (bottom). **(b)** Schematic illustration of the lateral confinement potential well created by patterned Cs atoms on the InSb(110) surface. Equipotential lines are shown in red. **(c)** Schematic of an anisotropic potential well resulting from an anisotropic patterning of Cs atoms. **(d)** Schematic energy level structure of QD states considering left: isotropic confinement ($\omega_x = \omega_y = \omega_0$), right: anisotropic confinement ($\omega_x < \omega_y$). **(e)** Diagram showing the multiplet structure under the influence of the different types of SOC terms, considering an isotropic confinement potential at $B = 0$ T. SOC leads to zero-field splittings in the multiplet. **(f)** Schematic magnetic field evolution of a given multiplet structure for states with different total magnetic moment $l+s$.

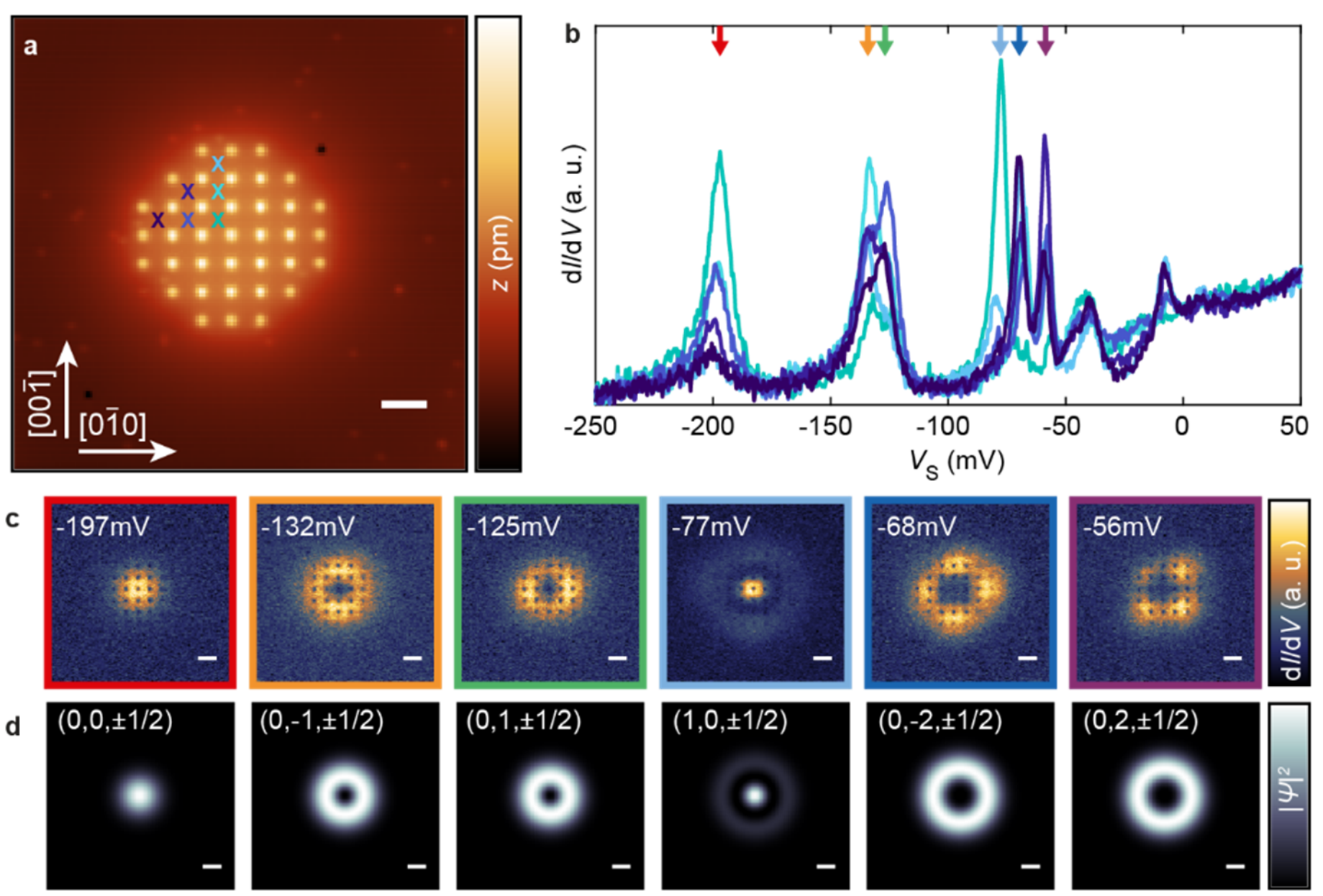

**Fig 2 | Atomic-scale quantum dot derived from 37 Cs atoms on InSb(110).**

**(a)** Constant-current STM image of an isotropic QD constructed from 37 Cs atoms ($V_S$ = 300 mV, $I_t$ = 5 pA, $\Delta z$ = 434 pm). **(b)** d$I$/d$V$ point spectroscopy measured at the six sites indicated with 'x' in (a) ($V_{stab}$ = 100 mV, $I_{stab}$ = 50 pA, $V_{mod,rms}$ = 0.5 mV). **(c)** d$I$/d$V$ maps at various bias voltages $V_S$ simultaneously recorded with the images in (a). The respective states imaged are color-coded and linked to the colored arrows in (b) ($z_{offset}$ = -100 pm, $V_{mod,rms}$ = 2 mV). **(d)** Calculated wavefunctions for the labeled eigenstates in a 2D quantum harmonic oscillator with quantum numbers ($n$,$l$,$s$) indicated, neglecting SOC. All scale bars: 5 nm.

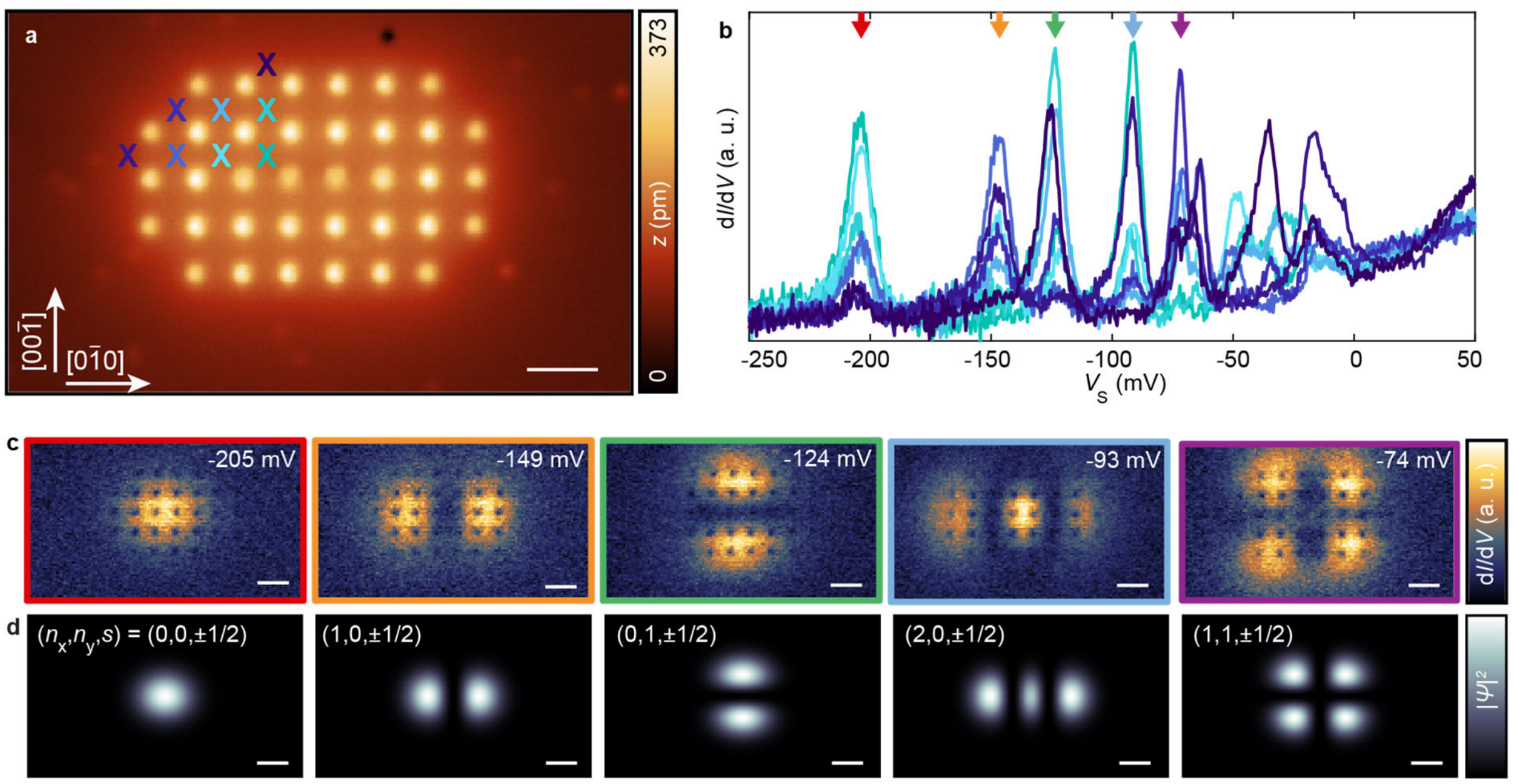

**Fig 3 Anisotropic quantum dot from 36 Cs atoms on InSb(110).**

**(a)** Constant-current STM image of an anisotropic QD constructed from 36 Cs atoms ($V_S$ = 300 mV, $I_t$ = 1 pA, $\Delta z$ = 373 pm). **(b)** d$I$/d$V$ point spectroscopy measured at the six sites indicated with 'x' in (a) ($V_{stab}$ = 50mV, $I_{stab}$ = 20 pA, $V_{mod,rms}$ = 0.5 mV)- **(c)** d$I$/d$V$ maps of the area in (a) taken at various bias voltages $V_S$. The respective states imaged are color-coded and linked to the colored arrows in (b) ($V_S$ = 300 mV, $I_t$ = 5 pA, $z_{offset}$ = -100 pm, $V_{mod,rms}$ = 2 mV). **(d)** Calculated wavefunctions for different eigenstates in an anisotropic 2D quantum harmonic oscillator ($\hbar\omega_x$ = 57 meV, $\hbar\omega_y$ = 81 meV), neglecting SOC. All scale bars: 5 nm.

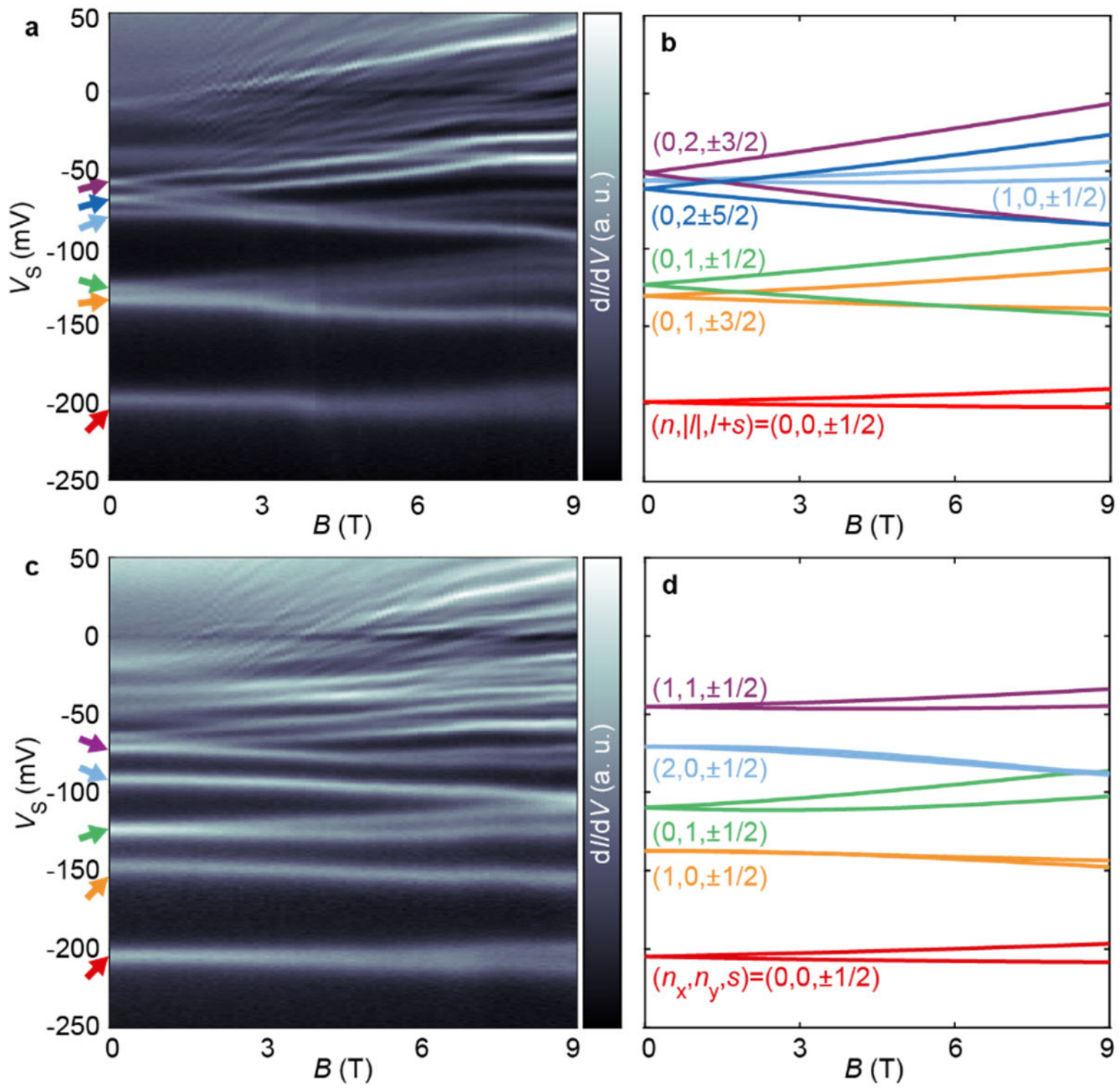

**Fig 4 | Quantifying the multiplet structure and computing the SOC contributions to the magnetic-field response.**

**(a)** d*I*/d*V* spectroscopy of the isotropic QD measured as a function of applied out-of-plane magnetic field. Spectra are averaged over the six indicated sites in Fig. 2a,b ($\Delta B$ = 100 mT, $V_{stab}$ = 100 mV, $I_{stab}$ = 50 pA, $V_{mod,rms}$ = 0.5 mV). **(b)** Calculated energy spectrum of an isotropic harmonic potential as a function of out-of-plane magnetic field using the effective Hamiltonian derived from the eight-band Kane model. The colors indicate the states shown in (a) at $B = 0$ T ($\kappa_x = \kappa_y = 1.06\times10^{15}$ eV/m$^2$, $v = 5.4\times10^7$ eV/m). **(c)** d*I*/d*V* spectroscopy of the anisotropic QD measured as a function of applied out-of-plane magnetic field. Spectra are averaged over the same eight sites as in Fig. 3a,b ($\Delta B$ = 100 mT, $V_{stab}$ = 50 mV, $I_{stab}$ = 20 pA, $V_{mod,rms}$ = 0.5 mV). **(d)** Calculated energy spectrum of an anisotropic harmonic potential as a function of out-of-plane magnetic field using the effective Hamiltonian obtained from the eight-band Kane model. The colors indicate the states shown in (b) at $B = 0$ T ($\kappa_x = 0.94\times10^{15}$ eV/m$^2$, $\kappa_y = 1.85\times10^{15}$ eV/m$^2$ $v = 5.1\times10^7$ eV/m).

Supplementary material for

# Spin-orbit coupling by design in quantum state engineering of atomically defined quantum dots

**Authors:**

Hermann Osterhage[1]†, Julian H. Strik[1]†, Ivan Ado[1], Anna M. H. Krieg[1], Daniel Wegner[2], Mikhail Titov[1], Alexander A. Khajetoorians[1*]

**S1 – Model Hamiltonian**

In the model Hamiltonian, we approximate the confinement potential created by the adsorbed Cs atoms as a sum of a triangular potential in $z$, perpendicular to the InSb(110) surface (*1*) (Fig. S1), and a parabolic potential in $x$ and $y$ (Fig. S2):

$$V = V_{\mathrm{z}} + V_{\mathrm{x,y}} = vz + \kappa_{\mathrm{x}} x^2 + \kappa_{\mathrm{y}} y^2. \qquad \text{S1}$$

Thus, the model Hamiltonian for conduction band electrons in zero magnetic field and without SOC can be separated into a vertical and a lateral component and reads

$$H_{00} = \frac{\hbar^2}{2m^*}\boldsymbol{\nabla}^2 + V = H_{\perp} + H_{00,\parallel} \qquad \text{S2}$$

with

$$H_{\perp} = -\frac{\hbar^2}{2m^*}\frac{\partial^2}{\partial z^2} + vz \qquad \text{S3}$$

$$H_{00,\parallel} = -\frac{\hbar^2}{2m^*}\left(\frac{\partial^2}{\partial x^2} + \frac{\partial^2}{\partial y^2}\right) + \kappa_{\mathrm{x}} x^2 + \kappa_{\mathrm{y}} y^2. \qquad \text{S4}$$

Solutions of the eigenvalue problem $H_{00}\Psi = E_{00}\Psi = (E_{\perp} + E_{00,\parallel})\Psi$ can be factorized as $\Psi = \Psi_{\perp}(z)\Psi_{00,\parallel}(x,y)$. $\Psi_{\perp}$ is given by the Airy functions with eigenvalues approximated by (*2*)

$$E_{\perp}(i) \approx \left(\frac{\hbar^2}{2m^*}\right)^{\frac{1}{3}}\left(\frac{3\pi v}{2}\left(i + \frac{3}{4}\right)\right)^{\frac{2}{3}}, i = 0,1,2,\ldots \qquad \text{S5}$$

and $E_{00,\parallel}$ are the levels of a quantum harmonic oscillator

$$E_{00,\parallel}(n_{\mathrm{x}}, n_{\mathrm{y}}) = \left(n_{\mathrm{x}} + \frac{1}{2}\right)\hbar\omega_{\mathrm{x}} + \left(n_y + \frac{1}{2}\right)\hbar\omega_{\mathrm{y}},\ \ n_i = 0,1,2,\ldots \qquad \text{S6}$$

with quantum numbers ($n_{\mathrm{x}}$,$n_{\mathrm{y}}$) and eigenfrequencies $\omega_{\mathrm{i}} = \sqrt{2\kappa_{\mathrm{i}}/m^*}$.

In the special case of isotropic confinement $(\omega_{\mathrm{x}} = \omega_{\mathrm{y}} = \omega_0)$, $H_{00,\parallel}$ has circular symmetry and we can write the eigenenergies and eigenstates as functions of the radial and angular variables $r$ and $\phi$:

$$E_{00,\parallel}(n,l) = (2n + |l| + 1)\hbar\omega_0 \tag{S7}$$

$$\Psi_{00,\parallel}^{(n,l)}(r,\phi) = e^{il\phi}\sqrt{\frac{m^*\omega_0}{\pi\hbar}}\sqrt{\frac{n!}{(n+|l|)!}}\,e^{-\frac{m^*\omega_0}{2\hbar}r^2}\left(\frac{r\sqrt{m^*\omega_0}}{\sqrt{\hbar}}\right)^{|l|} L_n^{|l|}\left(\frac{m^*\omega_0}{\hbar}r^2\right) \tag{S8}$$

with radial quantum number $n = 0,1,2,\ldots$ and angular quantum momentum number $l = 0, \pm 1, \pm 2, \ldots$. Here, $L_n^{|l|}$ are the generalized Laguerre polynomials.

The validity of the harmonic approximation for the lateral confinement is illustrated in Fig. S3. The confinement potential created by 36 ionic Cs atoms arranged in the shape of the quantum dot in Fig. 2 of the main paper is shown in Fig. S3a. Here, every Cs site is the center of a Yukawa potential with charge $Z = 0.8e$, dielectric constant $\varepsilon = 8.9$, and screening length $\lambda = 10$ nm as described in Ref. (*3*). A lateral cross section of the resultant confinement potential is shown in Fig. S3b. The parabolic fit shows that the sum of the Yukawa potentials created by each individual Cs atom can be approximated by a harmonic potential within the atomic quantum dot and is anharmonic outside of it.

Let us now outline a derivation of the full effective energy-dependent conduction band Hamiltonian that accounts for an out-of-plane magnetic field and for spin-orbit coupling (SOC). We start from the 8-band Kane model for semiconductors with the zinc-blende symmetry (*4*). The full (Kane) Hamiltonian $H^{(K)}$ in this model is an 8 by 8 operator-valued matrix. It is given by Eq. (16) of Ref. (*5*). Our goal is to reduce the full Schrödinger equation $H^{(K)}\Psi = E\Psi$ to an effective one in the conduction band. For this, we need to decouple the latter from the valence bands. In the presence of a confinement and a magnetic field, such a decoupling can only be done perturbatively.

In theories of semiconductor interfaces and nanostructures, effective Hamiltonians explicitly depend on the energy $E$ which effectively increases (or decreases) the gap energy (*6, 7*). However, such Hamiltonians have slightly different functional forms as compared to those of their bulk counterparts, for which, moreover, the energy dependence is usually disregarded. Nevertheless, one can modify the approach used to derive effective Hamiltonians in the bulk in such a way that the "bulk" functional form is retained, but the energy-dependence is taken into account(*8*). For the Kane Hamiltonian $H^{(K)}$, this procedure is outlined below.

In the bulk, the decoupling of the bands for $H^{(K)}$ can be performed by employing a unitary transformation of the form of Eq. (20) of Ref. (*5*). To the sub-leading order with respect to the inverse energy gap, $w$ in that equation is provided by Eq. (24) of the same work. To account for finite $E$, one should apply a similar unitary transformation to $H^{(K)} - E$ instead of $H^{(K)}$. This modification does not make any difference if all orders of the expansions are computed. Indeed, a unitary transformation does not modify a constant. However, effective theories under consideration are perturbative and include only a few expansion terms. Moreover, the expansion parameter is an energy difference between the bands which is affected by finite $E$.

To account for this, we replace the s-electrons Hamiltonian $H_s$ and the bandgap Hamiltonian $H_g$ in Eq. (24) of Ref. (*5*) by $H_s - E$ and $H_g - E$, respectively (note that now $H^{(K)} - E$ is considered instead of $H^{(K)}$). This leads to the same result for the effective (decoupled) conduction band Hamiltonian $H^{(c)}$ as in Eq. (31) of that paper, but with modified parameters. Below, we use the notation $H$ instead of $H^{(c)}$ for simplicity. Setting the energy origin at the conduction band edge ($i.e.,E_s = 0$) and disregarding the $\boldsymbol{B}$-dependent terms quadratic in $\boldsymbol{k}$ since they are negligible in our setup, we obtain (*4, 5*):

$$H = \frac{\hbar^2}{2m^*}\boldsymbol{k}^2 + V + H_{\mathrm{SOC}} + H_{\mathrm{Z}},$$

$$H_{\mathrm{SOC}} = \lambda_{\mathrm{SOC}}[\boldsymbol{\nabla} V \times \boldsymbol{k}]\boldsymbol{\sigma},\ H_{\mathrm{Z}} = \frac{g^*}{2}\mu_{\mathrm{B}}\boldsymbol{B}\boldsymbol{\sigma} \quad \text{S9}$$

where $\boldsymbol{k} = -i\boldsymbol{\nabla} - \frac{e}{\hbar}\boldsymbol{A}$ and the expressions

$$\frac{1}{m^*} = \frac{1}{m_e} + \frac{2\nu_1}{\hbar^2} + \frac{2\nu_2 E}{\hbar^2}, \qquad g^* = 2 + \frac{4m\lambda_1}{\hbar^2} + \frac{4m\lambda_2 E}{\hbar^2}, \qquad \lambda_{\mathrm{SOC}} = \lambda_2 + \lambda' = \lambda_2 + \frac{2\lambda_1}{E_{\mathrm{g}} + \Delta/3} \quad \text{S10}$$

$$\nu_{\mathrm{n}} = \frac{P^2}{3}\left(\frac{1}{\left(E_{\mathrm{g}} + \Delta + E\right)^n} + \frac{2}{\left(E_{\mathrm{g}} + E\right)^n}\right), \qquad \lambda_{\mathrm{n}} = \frac{P^2}{3}\left(\frac{1}{\left(E_{\mathrm{g}} + \Delta + E\right)^{\mathrm{n}}} - \frac{1}{\left(E_{\mathrm{g}} + E\right)^n}\right), \quad \text{S11}$$

determine the effective mass $m^*$, electron g-factor $g^*$, and spin-orbit coupling strength $\lambda_{\mathrm{SOC}}$ with material specific parameters $P$, $E_{\mathrm{g}}$, and $\Delta$.

We note that Eqs. (S9-S11) can be obtained using an alternative approach that closely resembles the usual method of deriving effective Hamiltonians for semiconductor interfaces and nanostructures(*8*). The correct (updated) expressions for the prefactors in front of the $\boldsymbol{B}$-dependent terms quadratic in $\boldsymbol{k}$ in the conduction band Hamiltonian are provided in (*9*). We also note that in this work we neglect the Darwin term $\Delta V$ for $V_{\mathrm{x,y}}$. It may weakly distort the effective band gap energy, but most likely does not contribute to spin splitting. Correct expression for the prefactor in front of the Darwin term in the effective conduction band Hamiltonian of the Kane model is yet unknown. The currently known expression, in our case, leads only to a negligible energy shift.

The effect of the energy $E$ on the model parameters is shown in Fig. S2: The confinement causes an additional offset in energy between the valence band and states in the conduction band, effectively modulating $m^*$, $g^*$, and $\lambda_{\mathrm{SOC}}$. The magnetic field enters in Eq. S9 through the Zeeman energy and through the electron momentum $\boldsymbol{k} = -i\boldsymbol{\nabla} - \frac{e}{\hbar}\boldsymbol{A}$, with the vector potential $\boldsymbol{A}$. In the symmetric gauge $\boldsymbol{A} = \frac{1}{2}\boldsymbol{B} \times \boldsymbol{r}$, with $\boldsymbol{B} = B\boldsymbol{e}_{\mathbf{z}}$ and by inserting Eq. S1 into Eq. S9 we get

$$\begin{aligned}H = &-\frac{\hbar^2\boldsymbol{\nabla}^2}{2m^*} - \frac{\hbar^2}{2m^*}\left(\frac{e}{2\hbar}B\left(x\boldsymbol{e}_{\mathrm{y}} - y\boldsymbol{e}_{\mathrm{x}}\right)\right)^2 + \frac{i\hbar}{2m^*}eB\left(x\frac{\partial}{\partial y} - y\frac{\partial}{\partial x}\right) \\ &+ vz + \kappa_{\mathrm{x}}x^2 + \kappa_{\mathrm{y}}y^2 + \frac{g^*}{2}\mu_{\mathrm{B}}\boldsymbol{B}\boldsymbol{\sigma} \\ &+ \lambda_{SOC}v[-i\boldsymbol{\nabla} \times \boldsymbol{\sigma}]_z + \lambda_{SOC}v\left[-\frac{e}{2\hbar}B\left(x\boldsymbol{e}_{\mathrm{y}} - y\boldsymbol{e}_{\mathrm{x}}\right) \times \boldsymbol{\sigma}\right]_z \\ &+ 2\lambda_{SOC}\left[\left(x\kappa_{\mathrm{x}}\boldsymbol{e}_{\mathrm{x}} + y\kappa_{\mathrm{y}}\boldsymbol{e}_{\mathrm{y}}\right) \times \left(-\frac{e}{2\hbar}B\left(x\boldsymbol{e}_{\mathrm{y}} - y\boldsymbol{e}_{x}\right)\right)\right] \cdot \boldsymbol{\sigma} \\ &+ 2\lambda_{SOC}\left[\left(x\kappa_{\mathrm{x}}\boldsymbol{e}_{\mathrm{x}} + y\kappa_{\mathrm{y}}\boldsymbol{e}_{\mathrm{y}}\right) \times (-i\boldsymbol{\nabla})\right] \cdot \boldsymbol{\sigma}\end{aligned} \quad \text{S12}$$

The first line in Eq. S12 is the kinetic term including effects from the magnetic vector field. The second line represents the confinement potential and Zeeman energy. And the other terms describe the influence of the SOC in dependence of the magnetic field and confinement strength. We will treat the SOC as a perturbation to the Hamiltonian without SOC: $H = H_0 + \delta H$. The wavefunctions of $\Psi = \Psi_{\perp}(z)\Psi_{0,\parallel}(x,y)$ And we define $H_0 = H_{\perp} + H_{0,\parallel}$, with

$$H_{0,\parallel} = -\frac{\hbar^2}{2m^*}\left(\frac{\partial^2}{\partial x^2}+\frac{\partial^2}{\partial y^2}\right) - \frac{\hbar^2}{2m^*}\left(\frac{e}{2\hbar}B\left(x\boldsymbol{e}_{\mathrm{y}} - y\boldsymbol{e}_{\mathrm{x}}\right)\right)^2 + \frac{i\hbar}{2m^*}eB\,\left(x\frac{\partial}{\partial y} - y\frac{\partial}{\partial x}\right) + \kappa_{\mathrm{x}}x^2 + \kappa_{\mathrm{y}}y^2 + \frac{g^*}{2}\mu_{\mathrm{B}}\boldsymbol{B\sigma},$$

including effects of the magnetic field but no SOC. The eigenvalues of the out-of-plane component remain the same as in Eq. S5. In the following we will treat the cases of isotropic and anisotropic lateral confinement potentials separately.

In the isotropic case $\left(\omega_{\mathrm{x}} = \omega_{\mathrm{y}} = \omega_0\right)$, the eigenstates of $H_{0,\parallel}$ are the well-known Fock-Darwin states (*10, 11*):

$$E_{0,\parallel}(n,l,s) = (2n + |l| + 1)\hbar\Omega - \frac{\hbar\omega_c}{2}l + g^*\mu_B Bs \qquad \text{S13}$$

$$\Psi_{0,\parallel}^{(n.l)} = \Psi_{n,l}(r,\phi,s) = e^{il\phi}\sqrt{\frac{m^*\Omega}{\pi\hbar}}\sqrt{\frac{n!}{(n+|l|)!}}\,e^{-\frac{m^*\Omega}{2\hbar}r^2}\left(\frac{r\sqrt{m^*\Omega}}{\sqrt{\hbar}}\right)^{|l|} L_n^{|l|}\left(\frac{m^*\Omega}{\hbar}r^2\right) \qquad \text{S14}$$

with the cyclotron frequency $\omega_c = -\frac{|e|B}{m^*}$, the spin projection $s$, and $\Omega = \sqrt{\omega_0^2 + \frac{\omega_c^2}{4}}$. Here we have used the fact that $l$ is an eigenvalue of the angular momentum operator $\boldsymbol{L_z} = -i\hbar(x\frac{\partial}{\partial y} - y\frac{\partial}{\partial x})$. We first compute the eigenvalues of $H^{(0)}$ for the twelve states lowest in energy of the first subband in $z$: $E = E_{\perp}(0) + E_{0,\parallel}(n,l,s_{\mathrm{z}})$ for

$$(n,l,s) = \left(0,0,\frac{1}{2}\right),\left(0,0,-\frac{1}{2}\right),\left(0,1,\frac{1}{2}\right),\left(0,1,-\frac{1}{2}\right),\left(0,-1,\frac{1}{2}\right),\left(0,-1,-\frac{1}{2}\right),$$
$$\left(0,2,\frac{1}{2}\right),\left(0,2,-\frac{1}{2}\right),\left(0,-2,\frac{1}{2}\right),\left(0,-2,-\frac{1}{2}\right),\left(1,0,\frac{1}{2}\right),\left(1,0,-\frac{1}{2}\right).$$

This is done self-consistently, taking into account the energy dependence of $g^*$ and $m^*$. We then compute the matrix elements of $\delta H$ perturbatively given this 12-state basis. $\delta H$ consist of correction terms induced by the out-of-plane confinement (Rashba-like terms) and of terms stemming from the in-plane confinement:

$$\delta H = H_R + H_{lat} \qquad \text{S15}$$

with

$$H_{lat} = \frac{2\kappa\lambda_{SOC}}{\hbar}\boldsymbol{L_z}\sigma_z + \frac{2\kappa\lambda_{SOC}m_{\mathrm{e}}}{\hbar^2}\mu_B(x^2+y^2)B\sigma_{\mathrm{z}} \qquad \text{S16}$$

$$H_R = v\lambda_{SOC}\left(-i\frac{\partial}{\partial x}\sigma_{\mathrm{y}} + i\frac{\partial}{\partial y}\sigma_{\mathrm{x}}\right) - \frac{v\lambda_{SOC}m_{\mathrm{e}}}{\hbar^2}\mu_B B\left(x\sigma_{\mathrm{x}} + y\sigma_{\mathrm{y}}\right). \qquad \text{S17}$$

**Origin of zero-field splittings:**

We see that both terms in Eq. S15 contribute to splittings in zero magnetic field and both include B-dependent terms. We will exemplarily discuss the zero-field splitting of states (0,±1,±1/2) here. These are fourfold degenerate in the absence of SOC. Most obvious is the zero-field splitting due to the first term of Eq. S16, leading to a splitting of states with parallel and antiparallel orientations of angular and spin moment. Consequently, the states (0,-1,1/2) and (0,1,-1/2) are lifted while (0,1,1/2) and (0,-1,-1/2) are pushed down in energy by $2\kappa\lambda_{\mathrm{SOC}}$, directly proportional to the lateral confinement strength. Additionally, the Rashba term couples states of $\Delta l = \pm 1$ and opposite spin projection (*12*). To estimate the influence of the Rashba term on the zero-field splitting, we calculated the relevant matrix elements for all of the twelve considered states analytically, assuming the unperturbed energy levels $E_{0,\parallel}$. This term results in a downward shift in energy of the states (0,-1,1/2) and (0,1,-1/2) by $2m^*v^2\lambda_{\mathrm{SOC}}^2/\hbar^2$. In total, the SOC-induced zero-field splitting in the (0,±1,±1/2) multiplet thus amounts to $\Delta E_{B=0}^{0,1} = 4\kappa\lambda_{\mathrm{SOC}} + 2m^*v^2\lambda_{\mathrm{SOC}}^2/\hbar^2$. These estimates are obtained neglecting the energy dependence of the effective parameters.

**Model for anisotropic quantum dots**

For anisotropic lateral confinement $(\omega_{\mathrm{x}} \neq \omega_{\mathrm{y}})$, we additionally treat the term $\frac{\omega_{\mathrm{c}}}{2}\boldsymbol{L_z}$ perturbatively, so that the total perturbation to $H_0 = H_\perp + H_{0,\parallel}$ reads $\delta H = H_R + H_{lat} - \frac{\omega_{\mathrm{c}}}{2}\boldsymbol{L_z}$, with

$$H_{0,\parallel} = -\frac{\hbar^2}{2m^*}\left(\frac{\partial^2}{\partial x^2} + \frac{\partial^2}{\partial y^2}\right) - \frac{\hbar^2}{2m^*}\left(\frac{e}{2\hbar}B\left(x\boldsymbol{e_y} - y\boldsymbol{e_x}\right)\right)^2 + \kappa_{\mathrm{x}}x^2 + \kappa_{\mathrm{y}}y^2 + \frac{g^*}{2}\mu_{\mathrm{B}}\boldsymbol{B\sigma}.$$

We parametrize the unperturbed spectrum with cartesian quantum numbers:

$$E_{0,\parallel}(n_x, n_y, s) = \left(n_{\mathrm{x}} + \frac{1}{2}\right)\hbar\Omega_{\mathrm{x}} + \left(n_{\mathrm{y}} + \frac{1}{2}\right)\hbar\Omega_{\mathrm{y}} + g^*\mu_B B\sigma_{\mathrm{z}}, \qquad \text{S18}$$

with $\Omega_{\mathrm{x,y}} = \sqrt{\omega_{\mathrm{x,y}}^2 + \frac{\omega_{\mathrm{c}}^2}{4}}$. The corresponding eigenstates are

$$\Psi_{0,\parallel}^{(n_x,n_y)}(x, y, s) = \sqrt{\frac{1}{\pi 2^{n_x}\, n_x!\, 2^{n_x}\, n_y!}}\, H_{n_x}\left(\sqrt{\frac{m^*\Omega_x}{\hbar}}\,x\right) H_{n_y}\left(\sqrt{\frac{m^*\Omega_y}{\hbar}}\,y\right) e^{-\frac{m^*}{2\hbar}(\Omega_{\mathrm{x}}x^2 + \Omega_{\mathrm{y}}y^2)} \qquad \text{S19}$$

with the Hermite polynomials $H_n$.We computed the unperturbed eigenenergies again in the ten states lowest in energy of the first subband in $z$ for

$$(n_x, n_y, s) = \left(0,0,\frac{1}{2}\right), \left(0,0,-\frac{1}{2}\right), \left(1,0,\frac{1}{2}\right), \left(1,0,-\frac{1}{2}\right), \left(0,1,\frac{1}{2}\right), \left(0,1,-\frac{1}{2}\right),$$
$$\left(2,0,\frac{1}{2}\right), \left(2,0,-\frac{1}{2}\right), \left(1,1,\frac{1}{2}\right), \left(1,1,-\frac{1}{2}\right).$$

**S2 - Determination of parameters**

The model Hamiltonian effectively contains two free parameters: the confinement strength perpendicular to the sample plane $v$ and the in-plane confinement strengths $\kappa_i$. In the following we describe how approximate values of these parameters are extracted from experimental STS data. In zero magnetic field, the ground state is doubly spin-degenerate. Neglecting corrections from SOC terms, these states split in magnetic field by $\Delta E_\mathrm{Z} = g^*(E)\mu_B B$, with an energy-dependent $g$-factor. We use the energy dependence of $g^*$ to extract an estimate of the vertical confinement strength. First, we quantify the experimentally measured splitting $\Delta E_\mathrm{Z}$ by fitting the states (0,0,±1/2) in Fig. 4a and c with two Lorentzian peaks. Then, we extract the total energy of the spin-degenerate ground state $E_\mathrm{GS} = E_\perp(0) + E_{0,\parallel}(0,0,s)$ at $B$ = 0 T through Eqs. S10 and S11 by a Zeeman fit to the experimentally measured splitting in magnetic field (see Figs. S4, S5). We find $g^*$ = 22.0±0.2 and $g^*$ = 22.5±0.1, respectively. Assuming a parabolic potential well, we take

$$E_{0,\parallel}(n = 0, l = 0, s) = E_{0,\parallel}(0,1,s) - E_{0,\parallel}(0,0,s) = \hbar\omega_0 \qquad \text{S20}$$

at $B$ = 0 T to be the energy separation between ground state and the next energy level for isotropic confinement $\hbar\omega_0 = 69$ meV (see Fig. S2). And likewise

$$E_{0,\parallel}(n_\mathrm{x} = 0, n_\mathrm{y} = 0, s) = \frac{1}{2}\left(E_{0,\parallel}(1,0,s) + E_\parallel^{(0)}(0,1,s)\right) - E_{0,\parallel}(0,0,s) = \frac{\hbar\omega_\mathrm{x} + \hbar\omega_\mathrm{y}}{2} \qquad \text{S21}$$

for anisotropic confinement, with $\hbar\omega_\mathrm{x} = 57$ meV and $\hbar\omega_\mathrm{y} = 81$ meV. This gives an estimate of the vertical confinement energy $E_\perp(0)$, and we calculate $v$ from Eq. S5. The resulting extracted parameters are $v = 5.4 \times 10^7$ eV/m for the isotropic case in Fig. 4a and $v = 5.1 \times 10^7$ eV/m for the anisotropic quantum dot in Fig. 4c.

To extract the in-plane confinement strength $\kappa$ for isotropic confinement, we fit the spatial decay of the density of states of the ground state at $B$ = 0 T with the squared wavefunction

$$\left|\Psi_{0,\parallel}^{(0,0)}(r)\right|^2 \propto e^{-\frac{m^*\omega_0}{\hbar}r^2}, \qquad \text{S22}$$

using the effective mass as a fitting parameter (see Fig. S6). Here, tunneling spectra were acquired at constant tip height at varying distance from the center of the quantum dot with stabilization parameters of $V_\mathrm{stab}$ = 100 mV, $I_\mathrm{stab}$ = 20 pA at the point closest to the center of the quantum dot, and $V_\mathrm{mod,rms}$ = 1 mV. The d$I$/d$V$ peak intensity of the ground state was fit using Eq. S22. This gives $\kappa = \frac{m^*\omega_0^2}{2}$ = 1.06×10$^{15}$ eV/m$^2$ for the case of isotropic confinement.

For the case of anisotropic confinement, we acquired d$I$/d$V$ spectra on vertical and horizontal lines through the quantum dot with 1 nm point spacing with stabilization conditions of $V_\mathrm{stab}$ = 100 mV, $I_\mathrm{stab}$ = 20 pA, $V_\mathrm{mod,rms}$ = 1 mV (see Fig. S7a). We fit the decay of

$$\left|\Psi_{0,\parallel}^{(0,0)}(x,y)\right|^2 \propto e^{-\frac{m_x^*\omega_x}{\hbar}x^2} e^{-\frac{m_y^*\omega_y}{\hbar}y^2} \qquad \text{S23}$$

separately, for the two lateral directions to the decay of the experimentally measured d$I$/d$V$ peak intensity normalized by the current (Fig. S7b). This gives two distinct values $\kappa_x = \frac{m_x{}^*\omega_x^2}{2}$ = 0.94×10$^{15}$ eV/m$^2$, and $\kappa_y = \frac{m_y{}^*\omega_y^2}{2}$ = 1.85×10$^{15}$ eV/m$^2$.

**S3 – Supplementary data**

**Quantum dot with circular symmetry**

The atomic QDs presented in Figs. 2 and 3 of the main paper are built with an approximate square atomic layout of interatomic distances of 3.2 nm between Cs atoms, making the density inside the QDs homogeneous, but introducing a spatial anisotropy due to the 4-fold symmetry. In Fig. S8a we show the arrangement of 36 atoms with near-circular symmetry. Energy spectra taken on six representative points within the QD are depicted in Fig. S8b. Note that the width of peaks in this experiment is larger compared to the case in Fig. 2b, leading to an overlap of all states with $(n,l,s)$ = (0,±1,±1/2) (orange). The spatially resolved eigenstates of the QD are shown in Fig. S8c and exhibit circular symmetry. Nevertheless, there is a zero-field splitting in the multiplet $(n,l,s)$ = (0,±2,±1/2) (blue and purple), indicative of SOC effects. The evolution of the spatially averaged spectrum as a function of out-of-plane magnetic field is shown in Fig. S8d and resembles the spectrum of the 4-fold symmetric QD is Fig. 4a.

**Independence of potential anisotropy and substrate anisotropy**

To test whether the anisotropy of the twofold-symmetric InSb(110) substrate plays a role in the energy spectrum of QDs constructed on it rather than the anisotropy of atomically manipulated patterns, we created a QD of 36 Cs atoms, like the one in Fig. 3 but rotated by 90° with respect to the InSb(110) surface. This QD, elongated along $[00\bar{1}]$ is shown in Fig. S9a. Point spectra taken on eight points in a representative quadrant of the QD are depicted in Fig. S9b. The maps of eigenstates taken at the peak energies (Fig. S9c) reveal that the eigenstates and their respective LDOS distributions are rotated by 90° as well. The magnetic field evolution of the energy spectrum is shown in Fig. S9d. We see that the magnitude of the energy splitting in magnetic field for states with nodes along the $x$- and $y$-direction is reversed compared to Fig. 4d. Here, the state with a node along $y$, now the strongly confined direction, does not show an appreciable splitting in $B$, proving that the relative magnitude of energy splittings originates from an anisotropy in the atomically induced confinement potential, rather than from the anisotropy of the substrate.

**Strongly anisotropic quantum dot**

In order to investigate whether the link between anisotropic confinement and magnetic field-induced energy splittings as demonstrated in Fig. 4c,d is universal, we constructed a highly anisotropic quantum dot from 48 Cs atoms on InSb(110) (see Fig. S10a). STS taken on four representative points is shown in Fig. S10b. By mapping the density of states at the peak energies we find a ground state with no nodal lines at -184 mV and excited states with one (-161 mV), two (-131 mV), and three (-96 mV) nodes along the $x$-direction (see Fig. S10c). The first state exhibiting a node along $y$ lies at -85 mV. Fig. S10d shows the eigenfunctions of a parabolic confinement potential with $\hbar\omega_x = 23$ meV, $\hbar\omega_y = 100$ meV, and $m^* = 0.02\, m_\mathrm{e}$ for comparison to illustrate that the eigenstates of the atomic quantum dot represent a case of strong anisotropy ($\omega_x < \omega_y$). The magnetic field evolution of the spatially averaged energy spectrum is shown in Fig. S10e. For the states (0,0,±1/2) and (1,0,±1/2) we observe a splitting in magnetic field. For the state (2,0,±1/2) we observe a broadening in high field at $B > 7$ T, indicative of a weaker splitting. For the state (3,0,±1/2) we do not see an observable splitting in magnetic field. However, the state (0,1,±1/2) with a node in the strongly confined direction again exhibits a stronger splitting in $B$, proving the link between the anisotropy of the atomically defined confinement potential and magnetic field-induced energy splittings in the spectrum.

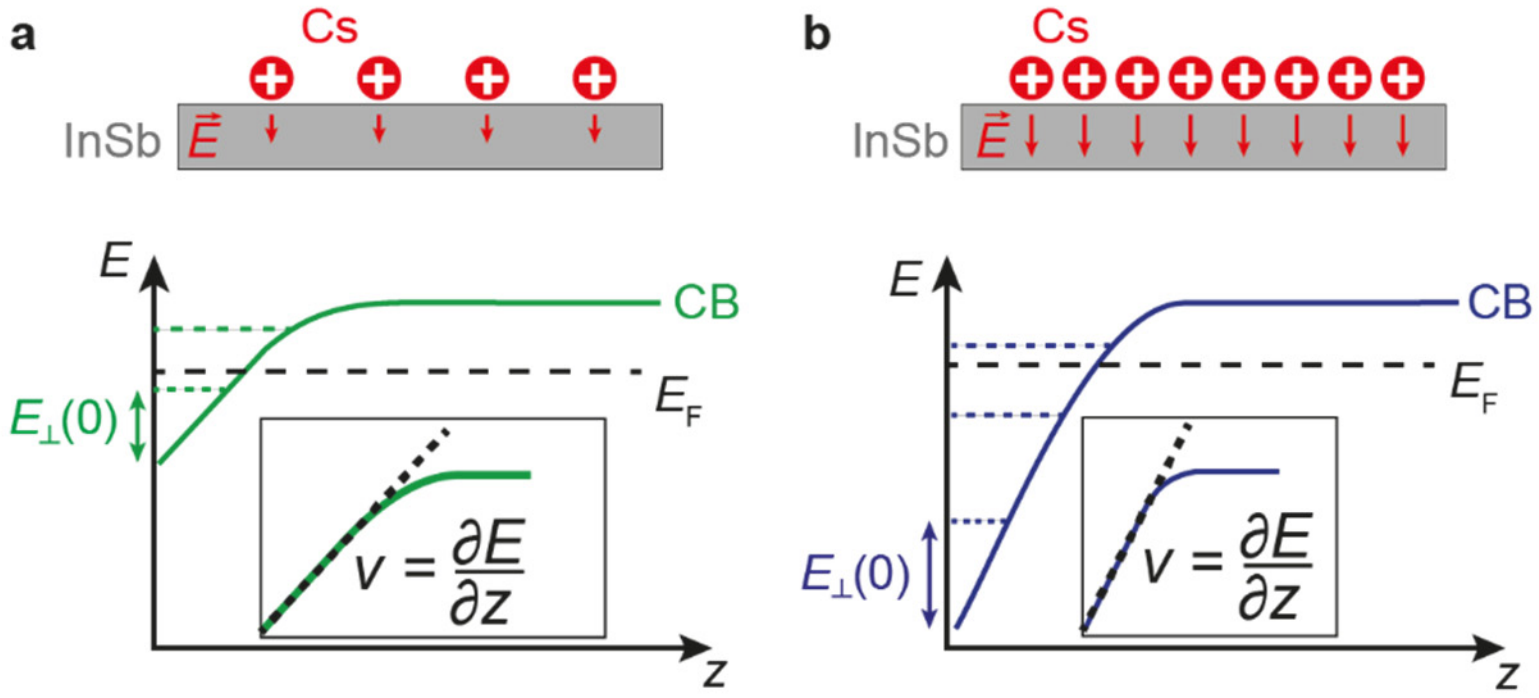

**Fig S1:** Band bending of the InSb conduction band induced by ionic Cs dopants on the (110) surface. (a) The electric field created by the dopants causes the conduction band to bend below the Fermi energy $E_F$ near the surface. Close to the surface, the band-bending profile is approximated by a triangular potential with gradient $v$. Broken lines symbolize energies of confinement states. (b) A higher Cs density leads to a stronger potential gradient and higher confinement energy.

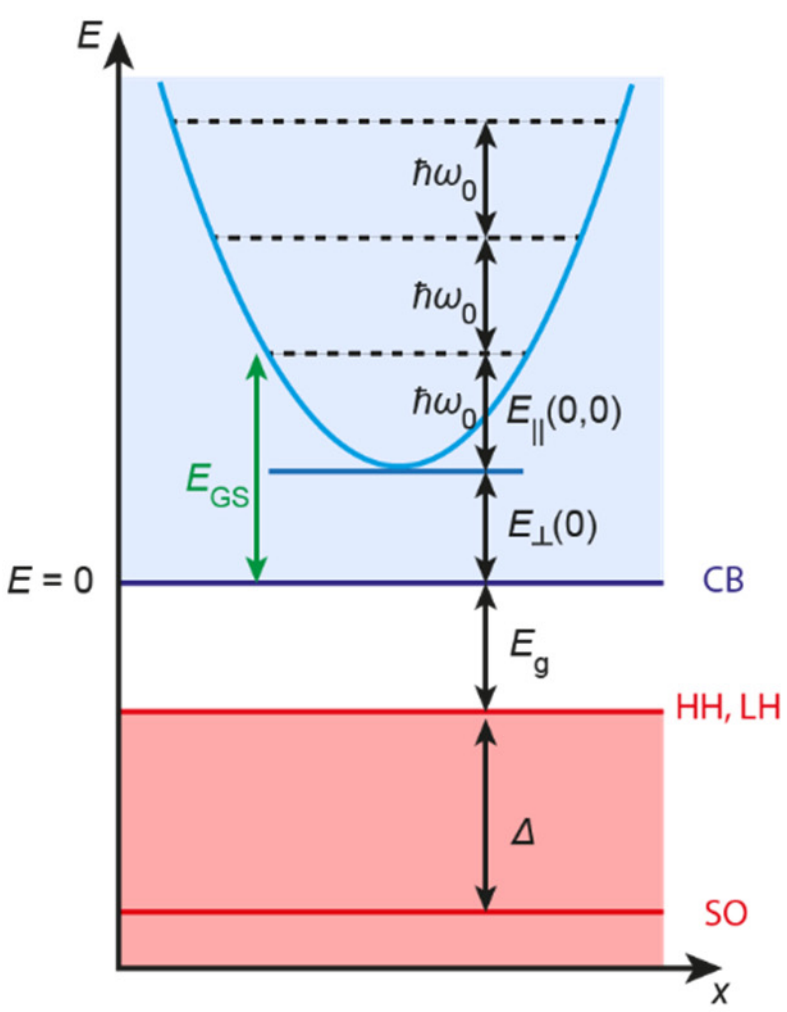

**Fig S2:** Schematic of band parameters and confinement energies. The band edges of the bulk conduction band (CB), the heavy-hole (HH) and light-hole (LH) valence bands as well as the split-off (SO) band are marked. Energies of states in the conduction band $E_\perp + E_{||}$ are defined with respect to the conduction band edge. The energy offset of the ground state $E_{GS}$ is marked in green.

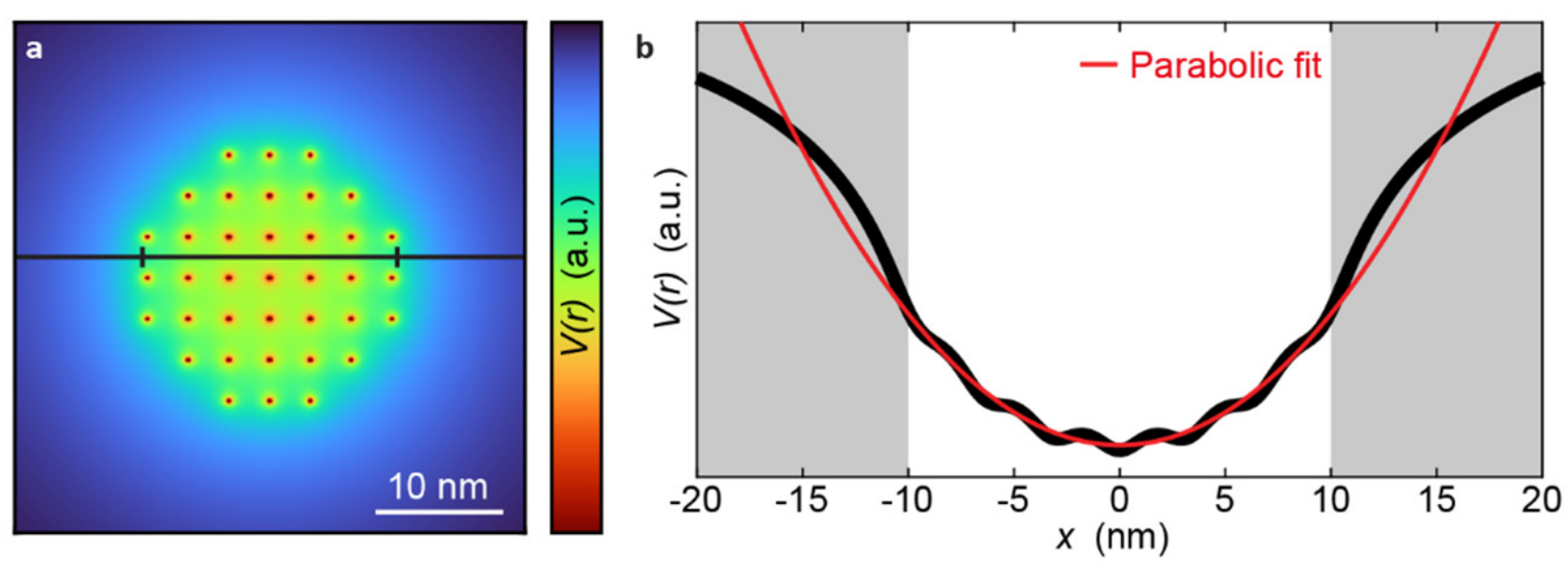

**Fig S3: (**a) Simulated potential landscape of an atomic quantum dot consisting of 37 Cs atoms on InSb(110) with a Yukawa potential ($Z = 0.8e$, dielectric constant $\varepsilon = 8.9$, and screening length $\lambda = 10$ nm) centered at each Cs site. (b) Cross section of the potential landscape along the black horizontal line. A parabolic fit to the potential inside the Cs pattern is shown in red.

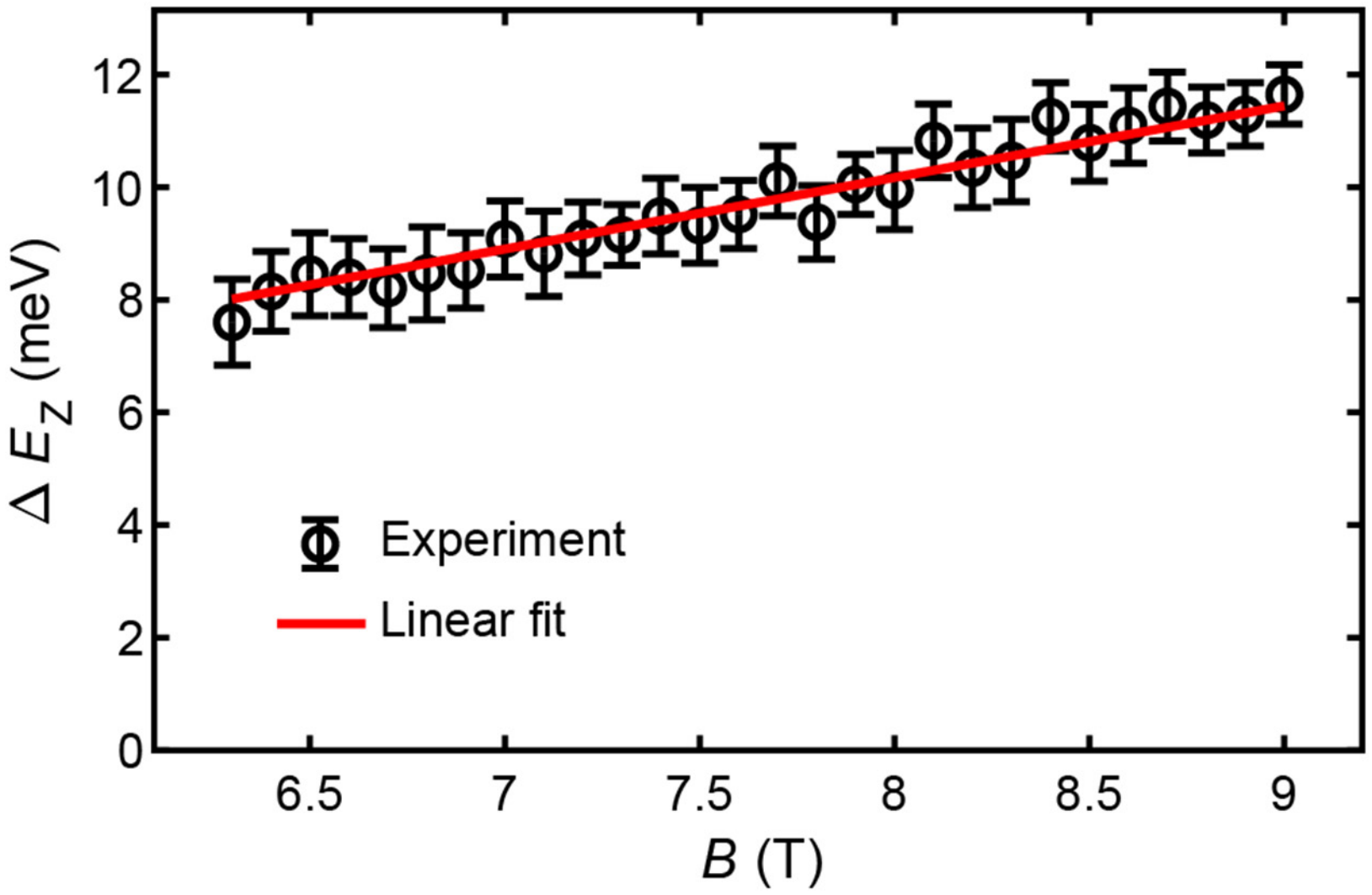

**Fig S4:** Energy splitting of the ground state (0,0,±1/2) in Fig. 4a of the main paper as a function of out-of-plane magnetic field with a linear Zeeman fit to the data. Error bars represent the uncertainty in the Zeeman splitting extracted from Lorentzian fits to the spectrum.

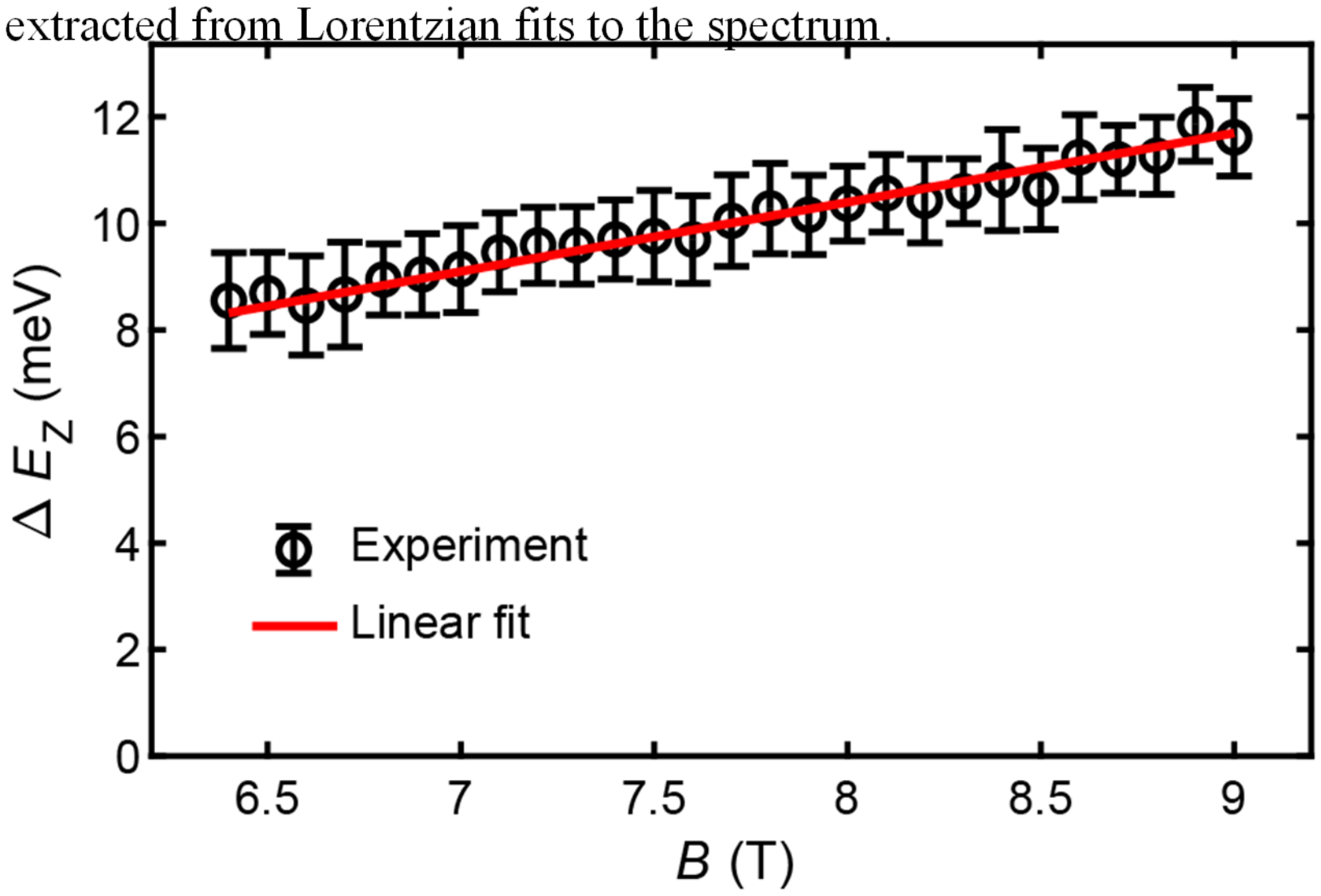

**Fig S5:** Energy splitting of the ground state (0,0,±1/2) in Fig. 4c of the main paper as a function of out-of-plane magnetic field with a linear Zeeman fit to the data. Error bars represent the uncertainty in the Zeeman splitting extracted from Lorentzian fits to the spectrum.

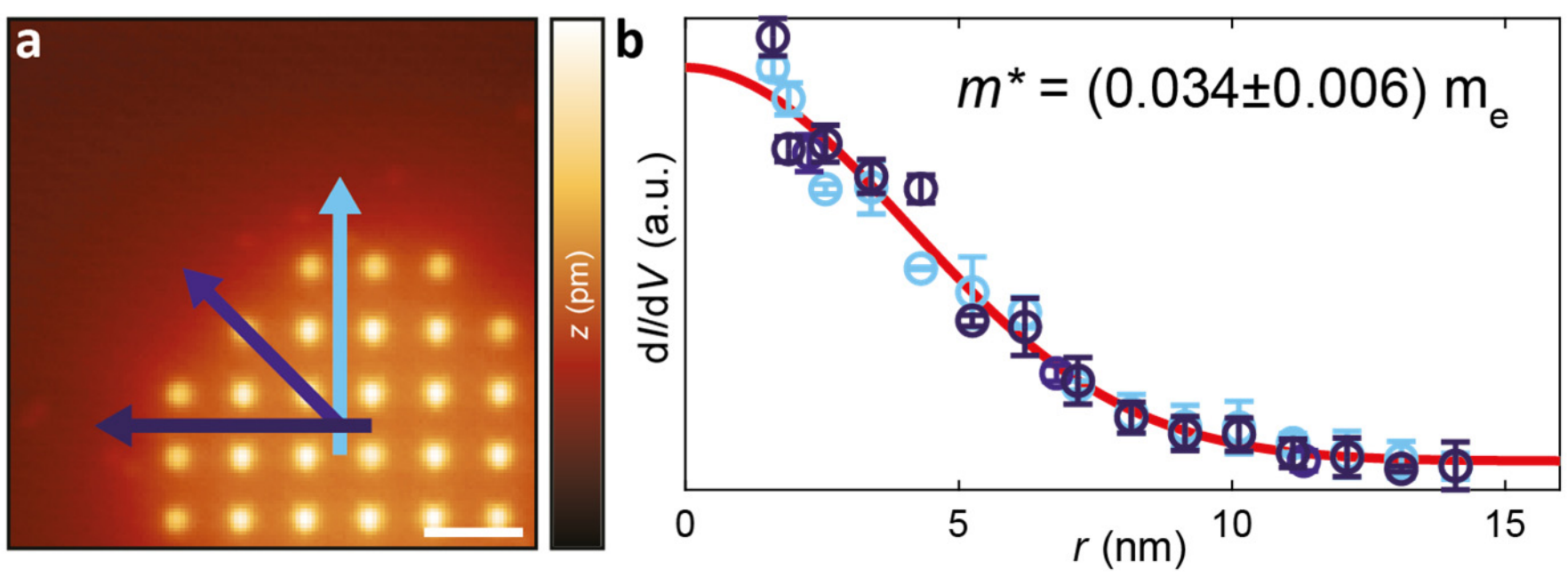

**Fig S6: a)** Schematic of lines along which tunneling spectroscopy was performed on the quantum dot shown in Fig. 2 of the main paper. Stabilization: $V_{stab}$ = 100 mV, $I_{stab}$ = 20 pA at the point closest to the center of the quantum dot, $V_{mod,rms}$ = 1 mV. **b)** Peak intensity of the ground state (0,0,±1/2) as a function of distance from the center of the quantum dot measured along the three directions marked in (a), representing the squared ground state wavefunction. Error bars represent the difference in intensity at the peak vs. 1 mV off of it as an upper bound of the uncertainty. The red line is a Gaussian fit to the data.

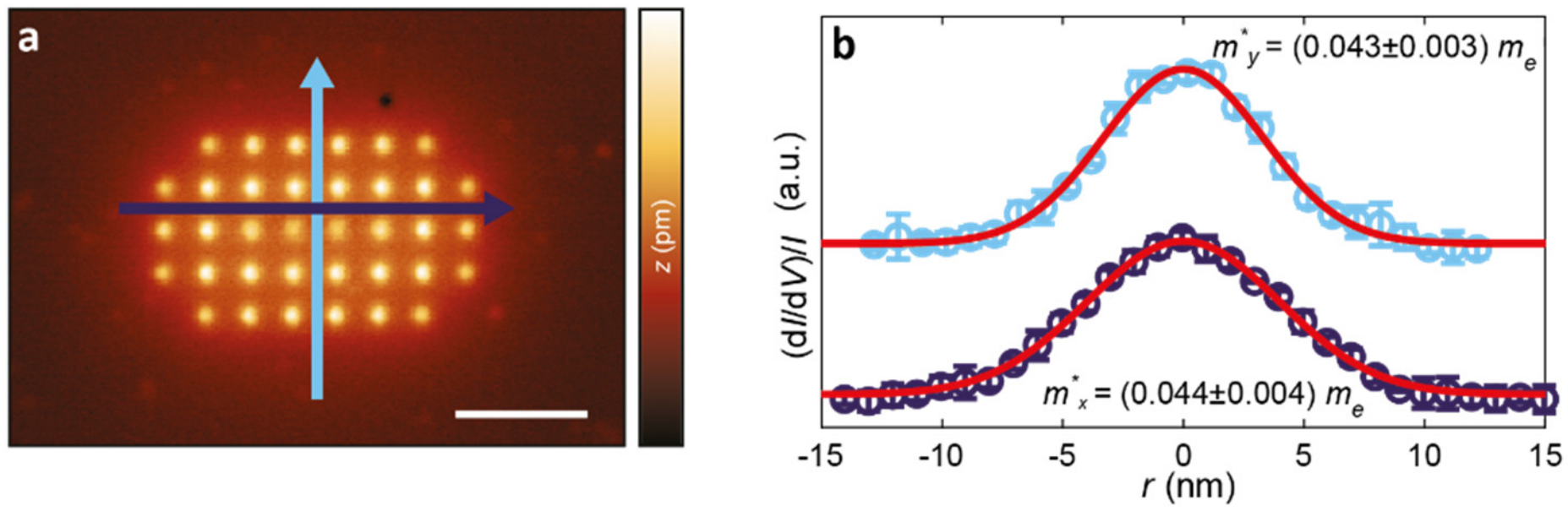

**Fig S7: a)** Schematic of lines along which tunneling spectroscopy was performed on the quantum dot shown in Fig. 3 of the main paper. Stabilization: $V_{stab}$ = 100 mV, $I_{stab}$ = 20 pA, $V_{mod,rms}$ = 1 mV. **b)** d*I*/d*V* peak intensity normalized by tunnel current *I* of the ground state (0,0,±1/2) ), representing the squared ground state wavefunction, as a function of distance from the center of the quantum. Dark (light) blue dots represent the data measured along the horizontal (vertical *y*) direction marked in (a). The data in light blue is artificially offset. Error bars represent the difference in intensity at the peak vs. 1 mV off of it as an upper bound of the uncertainty. The red lines represent separate Gaussian fits to the data for the *x* and *y* direction, representing the spatial decay of squared ground state wavefunction.

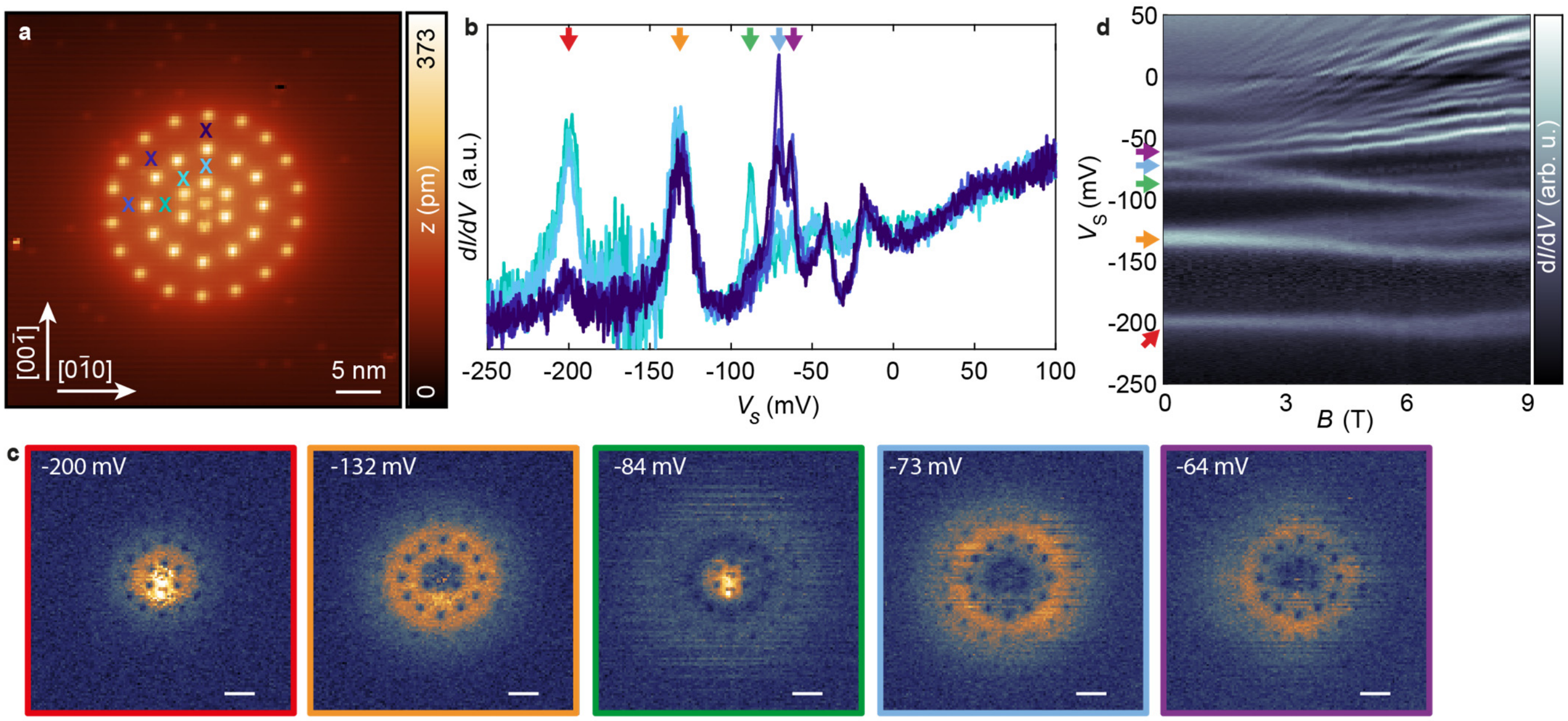

**Fig S8: Circular atomic-scale quantum dot constructed from 37 Cs atoms.**
**(a)** Constant-current STM image of a circular, isotropic QD constructed from 37 Cs atoms ($V_S$ = 300 mV, $I_t$ = 5 pA, $\Delta z$ = 373 pm). **(b)** d$I$/d$V$ point spectroscopy measured at the six sites indicated with 'x' in (a) ($V_{stab}$ = 100 mV, $I_{stab}$ = 50 pA, $V_{mod,rms}$ = 0.5 mV). **(c)** d$I$/d$V$ maps at various bias voltages $V_S$ simultaneously recorded with the images in (a). The respective states imaged are color-coded and linked to the colored arrows in (b) ($z_{offset}$ = -100 pm, $V_{mod,rms}$ = 2 mV). **(d)** d$I$/d$V$ spectroscopy of the isotropic QD measured as a function of applied out-of-plane magnetic field. Spectra are averaged over the six sites in (a) ($\Delta B$ = 100 mT, $V_{stab}$ = 100 mV, $I_{stab}$ = 50 pA, $V_{mod,rms}$ = 0.5 mV). All scale bars: 5 nm.

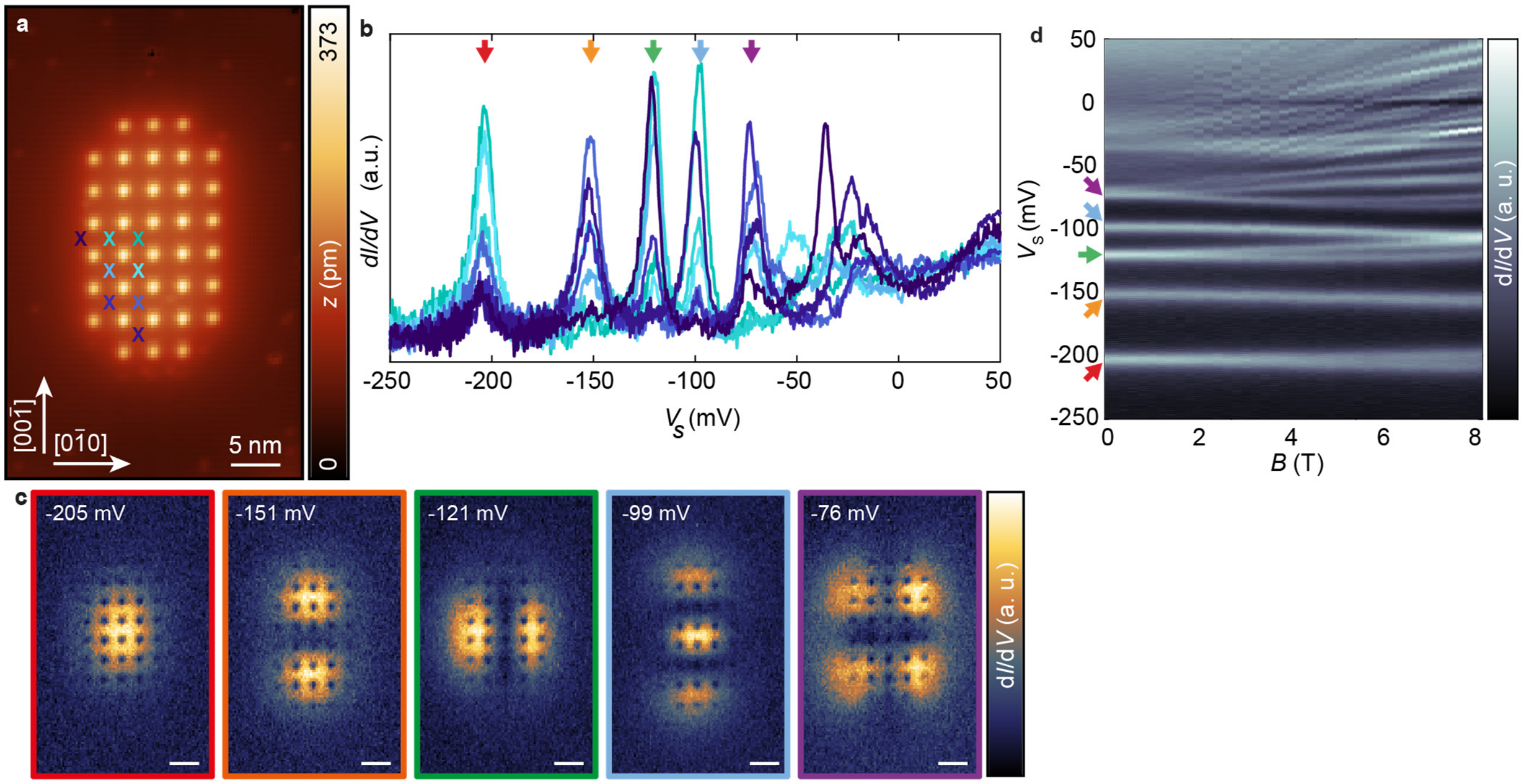

**Fig S9: Anisotropic quantum dots of 36 Cs atoms with strong quantization along [0-10].** Constant-current STM image of an anisotropic QD constructed from 36 Cs atoms ($V_S$ = 300 mV, $I_t$ = 5 pA, $\Delta z$ = 373 pm). **(b)** d$I$/d$V$ point spectroscopy measured at the eight sites indicated with 'x' in (a) ($V_{stab}$ = 50 mV, $I_{stab}$ = 20 pA, $V_{mod,rms}$ = 0.5 mV). **(c)** d$I$/d$V$ maps at various bias voltages $V_S$ simultaneously recorded with the images in (a). The respective states imaged are color-coded and linked to the colored arrows in (b) ($z_{offset}$ = -100 pm, $V_{mod,rms}$ = 2 mV). **(d)** d$I$/d$V$ spectroscopy of the isotropic QD measured as a function of applied out-of-plane magnetic field. Spectra are averaged over the six sites in (a) ($\Delta B$ = 100 mT, $V_{stab}$ = 50 mV, $I_{stab}$ = 20 pA, $V_{mod,rms}$ = 0.5 mV). All scale bars: 5 nm.

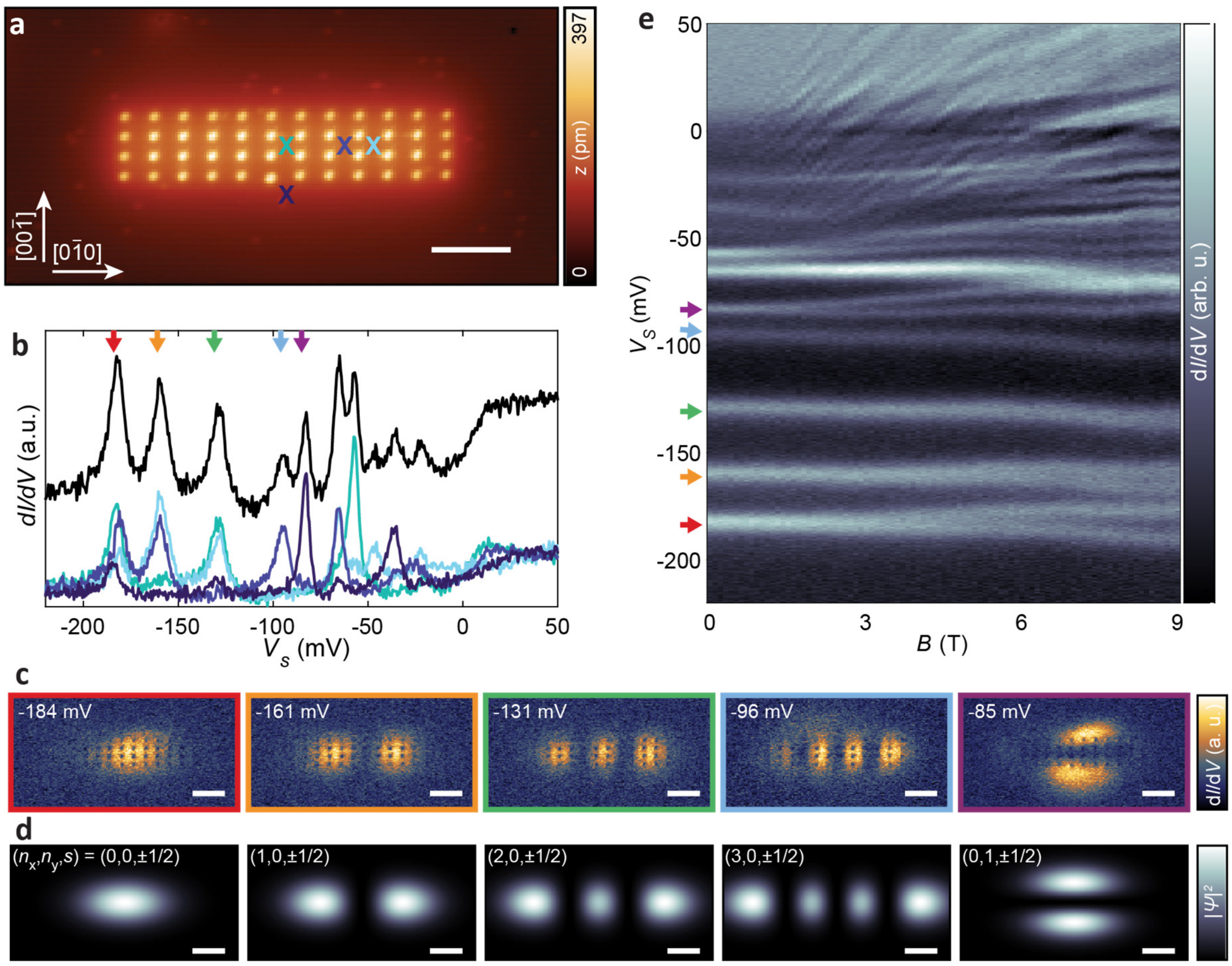

**Fig S10: Highly anisotropic quantum dot made of 48 Cs atoms. a)** Constant-current STM image of a highly anisotropic QD constructed from 48 Cs atoms ($V_S$ = 300 mV, $I_t$ = 5 pA, $\Delta z$ = 397 pm). **(b)** d$I$/d$V$ point spectroscopy measured at the four sites indicated with 'x' in (a). The averaged spectrum over the four sites is shown in black, vertically offset ($V_{stab}$ = 100 mV, $I_{stab}$ = 50 pA, $V_{mod,rms}$ = 0.7 mV). **(c)** d$I$/d$V$ maps at various bias voltages $V_S$ simultaneously recorded with the images in (a). The respective states imaged are color-coded and linked to the colored arrows in (b) ($z_{offset}$ = -100 pm, $V_{mod,rms}$ = 2 mV). **(d)** Calculated wavefunctions for the labeled eigenstates in a 2D quantum harmonic oscillator ($\hbar\omega_x$ = 23 meV, $\hbar\omega_y$ = 100 meV, $m_{eff}$ = 0.02 $m_e$). **(e)** d$I$/d$V$ spectroscopy of the isotropic QD measured as a function of applied out-of-plane magnetic field. Spectra are averaged over the six sites in (a) ($\Delta B$ = 100 mT, $V_{stab}$ = 100 mV, $I_{stab}$ = 50 pA, $V_{mod,rms}$ = 0.7 mV). All scale bars: 10 nm.